\documentclass[12pt,preprint]{aastex}
\usepackage{amsmath, amssymb}

\shortauthors{T.Hosokawa \& S.Inutsuka}
\shorttitle{Dynamical Formation of Dark Molecular Hydrogen Clouds}

\begin{document}

\title{Dynamical Formation of the Dark Molecular Hydrogen Clouds around 
Diffuse H~II Regions}
\author{Takashi Hosokawa\altaffilmark{1} and 
     Shu-ichiro Inutsuka\altaffilmark{2}}

\altaffiltext{1}{Division of Theoretical Astrophysics, 
National Astronomical Observatory, Mitaka, Tokyo 181-8588, Japan ; 
hosokawa@th.nao.ac.jp}
\altaffiltext{2}{Department of Physics, Kyoto University, Kyoto 606-8502 ;
inutsuka@tap.scphys.kyoto-u.ac.jp}

\begin{abstract}
We examine the triggering process of molecular cloud formation
around diffuse H~II regions. 
We calculate the time evolution of the shell as well as of the H~II region
in a two-phase neutral medium, solving the UV and FUV radiative
transfer, the thermal and chemical processes in the time-dependent 
hydrodynamics code. In the cold neutral medium, the ambient 
gas is swept up in the cold ($T \sim 30-40$~K) and dense 
($n \sim 10^3~{\rm cm}^{-3}$) shell around the HII region. 
In the shell, H$_2$ molecules are formed from the swept-up H~I gas, 
but CO molecules are hardly formed. This is due to the different 
efficiencies of the self-shielding effects between H$_2$ and CO molecules. 
The reformation of H$_2$ molecules is more efficient with a
higher-mass central star. The physical and chemical properties of 
gas in the shell are just intermediate between those of
the neutral medium and molecular clouds observed by CO emissions.
The dense shell with cold HI/H$_2$ gas easily becomes gravitationally 
unstable, and breaks up into small clouds. The cooling layer just 
behind the shock front also suffers from thermal instability, and 
will fragment into cloudlets with some translational motions.
We suggest that the predicted cold ``dark'' HI/H$_2$ gas  
should be detected as the H~I self-absorption (HISA) feature.
We have sought such features in recent observational data,
and found shell-like HISA features around the giant H~II regions, 
W4 and W5. The shell-like HISA feature shows good spatial correlation
with dust emission, but poor correlation with CO emission.
Our quantitative analysis shows that the HISA cloud
can be as cold as $T \sim $ a few $\times$ 10~K.
In the warm neutral medium, on the other hand, the expanding diffuse 
H~II region is much simpler owing to the small pressure excess. 
The UV photons only ionize the neutral medium and 
produce a warm ionized medium.  
\end{abstract}


\section{Introduction}

The UV $(h\nu \geq 13.6~{\rm eV})$ and FUV 
$(11~{\rm eV} \lesssim h\nu \leq 13.6~{\rm eV})$ radiations
from massive stars are important for the
cycle of the interstellar medium (ISM). 
Stellar UV radiation ionizes the neutral medium, and this
photoionization is the main heating process in the ionized medium.
The diffuse warm ionized medium (WIM) is widely distributed throughout 
our galaxy \citep{Hf03}, and the photoionization is a 
promising process to supply the ubiquitous WIM \citep[e.g.,][]{MC93, DS94}. 
The FUV radiation photodissociates the molecular medium, and
photoelectric heating via the dust absorption of FUV photons
is the primary heating process in the neutral medium.
Since the chemical and thermal balances determine the equilibrium  
of the neutral medium for a given density, the strength of the FUV 
radiation field inevitably determines the physical state of the neutral 
medium \citep[e.g.,][]{Wl95, Wl03}. 
The well-known equilibrium $n$-$P$ curve shows that
the two different phases can be achieved at some pressure levels.
It has been generally accepted that the observed cold neutral 
medium (CNM) and warm neutral medium (WNM) correspond to these two
thermally stable states \citep[see][for a recent review]{Cx05}. 
Besides its significance in the diffuse ISM, 
the UV radiation also works to form the cold and dense molecular
structure. The shock front (SF) sometimes emerges ahead of the 
ionization front (IF) owing to high pressure in the H~II region.
The compression of the neutral
medium behind the SF is one of the possible processes to form
the molecular cloud \citep[e.g.,][]{ki00, ki02, Bg04}.
The ambient medium is swept up by the expanding SF, and
molecules reform in the dense shell. 
The density and strength of FUV radiation in the shell
determine the formation efficiency of molecules \citep{El93}.
Some observations have shown clear features of this mode
of triggering molecular cloud formation 
\citep[e.g.,][]{Gir94, Ym03, Ym06}.  
However, there is still a missing link between the neutral medium 
and molecular clouds observed with the CO emission.
Even if significant H$_2$ molecules are formed in the shell,
the intermediate cold gas phase without CO molecules is 
observationally ``dark''.
Some recent studies have reported signs of such ``dark''
gas phases \citep{On01, Gr05}. 

In our previous papers 
\citep[][hereafter, Papers I and II]{HI05, HI06a}, we have studied
the expanding H~II regions in the molecular clouds focusing on
the detailed structure of the swept-up shell. 
We have shown that most of the swept-up gas remains in the shell
as cold and dense molecular gas, and star-formation triggering 
is expected with various central stars and ambient number densities.
After the destruction of the cloud, on the other hand, the stellar
UV/FUV radiation spreads to the diffuse neutral medium.
In this paper, we explore how the stellar UV/FUV radiation works
for the cycle of the ISM. We calculate the time evolution
of the shell as well as of the H~II region in the ambient
neutral medium. While the expanding H~II region photoionizes 
the neutral medium, the swept-up neutral medium may be recycled
into the molecular gas.
We calculate models with various ambient number densities
(CNM and WNM), and central stars to explore the role of the 
H~II regions expanding in the diffuse ISM.
We show that the cold ($T \sim 30-40$~K) and dense 
($n \sim 10^3~{\rm cm}^{-3}$) shell is sometimes formed 
around the H~II region, and H$_2$ molecules
accumulate there without producing any CO molecules.
This is just the intermediate gas phase between
the neutral medium and the molecular clouds. The CO emission
is not available to enable the search of this cold HI/H$_2$ gas.
Instead, we try to hunt the predicted ``dark'' hydrogen with
the HI self-absorption (HISA). We find the shell-like
HISA around the giant H~II regions, W4 and W5.  
Our quantitative analysis shows that this HISA cloud can be
the intermediate gas phase, just as predicted in our numerical 
calculations.

The organization of this paper is as follows:
In \S~\ref{sec:num}, we briefly review our numerical method.
The calculated models are also listed in this section.
In \S~\ref{sec:nres}, we present the results of the numerical
calculations. In \S~\ref{ssec:ex_cnm} and \ref{ssec:ex_wnm}, we 
examine the expansion of the H~II region in the ambient CNM and 
WNM respectively. The findings based on the observational data 
are shown in \S~\ref{sec:obs}. \S~\ref{sec:disc} and \ref{sec:conc} 
are assigned to the discussions and conclusions.

\section{Numerical Modeling}
\label{sec:num}

We solve the UV/FUV radiative transfer and thermal/chemical processes
in a time-dependent hydrodynamics code.
We have described our numerical code in detail in Paper II, 
and only mention some points here.
The numerical scheme of the hydrodynamics is based on the 
second-order Lagrangian Godunov method \citep[e.g.,][]{vL79}.
The frequency-dependent UV/FUV radiative transfer is solved with 
on-the-spot approximation.  We include 17 heating/cooling
processes in the energy equation; for example, H-photoionization heating, 
photoelectric heating, H-recombination cooling, and line cooling via 
Ly-$\alpha$, OI~(63.0$\mu$m, 63.1$\mu$m),
OII~(37.29$\mu$m), CII~(23.26$\mu$m, 157.7$\mu$m), rovibrational 
transitions of H$_2$ and CO molecules and so on.
The energy transfer by thermal conduction is also included
in the energy equation to resolve the Field length of the thermally
unstable media \citep[][]{ki04}. The minimal set of the chemical 
reactions is implicitly solved for e, H, H$^+$, H$_2$, C$^+$, and 
CO species. The simplified reaction scheme between C$^+$ and CO is 
adopted, following \citet{NL97}. Since the size distribution of grains 
is uncertain, we include the dust absorption only outside the H~II region.
We have confirmed that dust absorption in the H~II region 
is not important in the diffuse ambient medium, owing to the low 
column density of the H~II region \citep{Pt72,Ar04}.

We add the effect of the FUV background radiation field to
our code. The gas temperature and the ionization rate in the ambient
neutral medium are initially set at the equilibrium values for the 
given density and FUV background field normalized in Habing units,
$G_{\rm FUV}$ \citep{Hb68}. 
If a dense shell forms around the H~II region, molecular
gas may form from the swept-up neutral gas. 
However, the reformed molecules in the shell can be photodissociated
by FUV photons, both from the central star and from the
background field. The penetration of the FUV background radiation 
into the shell is treated by the analytic shielding function, measuring
the optical depth from the SF \citep{DB96, Le96}. 
Some recent codes for the steady PDR have included more detailed
calculation of the level populations and line transfers,
and shown that the resultant dissociation rates possibly deviate from 
those obtained by the approximate treatment of the shielding effect
\citep[e.g.,][]{Sw05, Lp06}.
For comparison, we solved the structure of the irradiated slab that 
resembles the shell appearing in our calculations, using the Meudon 
PDR code \citep{Lp06}. We confirmed that these accurate calculations 
also show the physical/chemical structure very similar to that of the 
shell explained below.

\begin{figure}[tb]
\begin{center}
\includegraphics[width=0.8\hsize]{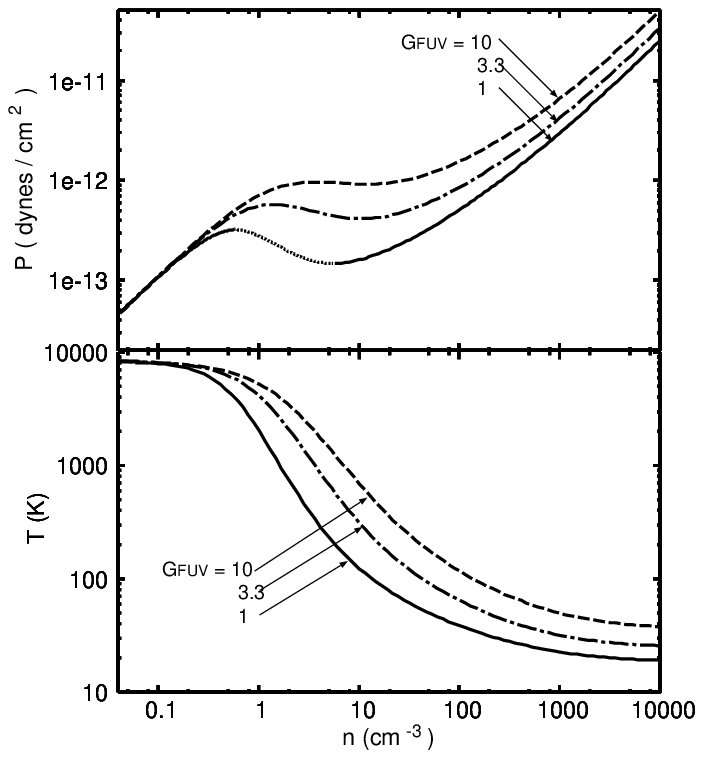}
\caption{Equilibrium gas temperature and pressure of the neutral
medium for a given gas density and FUV radiation field. 
In each panel, the broken, dotted, and solid lines 
respectively represent the equilibrium curves with different FUV
 radiation fields, $G_{\rm FUV} = 10$, 3.3, and 1.
The dotted curve in the upper panel denotes the thermally-unstable
equilibrium states for $G_{\rm FUV} = 1$.
 }
\label{fig:eqil}
\end{center}
\end{figure}

We adopt $G_{\rm FUV} = 1.0$ as the standard value of the FUV background
field, and include no attenuation in the ambient medium.
The equilibrium $n$-$P$ and $n$-$T$ curves for $G_{\rm FUV} = 1.0$ are
presented in Figure \ref{fig:eqil} \citep[also see][]{Wl95,ki00}.
The cooling rate exceeds the heating rate above the equilibrium
curve, and vice versa. The thermally unstable equilibrium states
for $G_{\rm FUV}=1.0$ are denoted by the dotted lines.
As the upper panel of Figure \ref{fig:eqil} shows, the two 
thermally stable equilibrium states can coexist under the same gas 
pressure of $1.5 \lesssim P \lesssim 3.2~(10^{-13} {\rm dynes/cm}^2)$, 
which are the CNM and WNM. The thermal balance is achieved mainly 
between the photoelectric heating and radiative cooling by [C~II] 
157.7$\mu$m (Ly-$\alpha$ and [O~I] $63.1\mu$m) in the CNM (WNM).
In Figure \ref{fig:eqil}, we also plot equilibrium curves with the
stronger FUV fields of $G_{\rm FUV} = 3.3$ and 10. 
The equilibrium temperature increases with the stronger FUV
field, owing to the efficient photoelectric heating.
In our numerical models presented in \S~\ref{sec:nres}, 
the FUV radiation field can be as high as 
$G_{\rm FUV} > 10$ just outside the H~II region, owing to the proximal 
massive star. As shown in \S~\ref{ssec:ex_cnm} below, the SF emerges and
a dense shell forms around the H~II region expanding in
the ambient CNM. These equilibrium curves are useful to interpret 
the calculated physical states in the shell. 
We analyze the transition of physical states across the SF
with these equilibrium curves in \S~\ref{ssec:ex_cnm}.  

\begin{table}[htb]
\label{tb:md_hi}
\begin{center}
Table 1. Model Parameters
\\[3mm]
\begin{tabular}{l|cccccc}
\hline
Model  & $M_*~(M_\odot)^a$ & 
         $n_{\rm H,0} ({\rm cm^{-3}})^b$ &
         $\log S_{\rm UV} ({\rm s}^{-1})^c$ 
       & $\log S_{\rm FUV} ({\rm s}^{-1})^d$ &
         $R_{\rm st}~({\rm pc})^e$ & $t_{\rm dyn}~({\rm Myr})^f$ \\
\hline
\hline
CNM-S64     & 63.8  & 10  & 49.35 & 49.16 & 18.16  & 1.67 \\
CNM-S41     & 40.9  & 10  & 48.78 & 48.76 & 11.73  & 1.08 \\
CNM-S22     & 21.9  & 10  & 48.10 & 48.33 & 6.96   & 0.64 \\
CNM-S18     & 17.5  & 10  & 47.02 & 47.99 & 3.04   & 0.28  \\
CNM-S12     & 11.7  & 10  & 45.38 & 47.39 & 0.86   & 0.08  \\
WNM-S41     & 40.9  & 0.1 & 48.78 & 48.76 & 252.7  & 23.27 \\
WNM-S18     & 17.5  & 0.1 & 47.02 & 47.99 & 65.45  & 6.03 \\
WNM-S12     & 11.7  & 0.1 & 45.38 & 47.39 & 18.59  & 1.71 \\
\hline                           
\end{tabular}
\noindent
\end{center} 
a : mass of the central star, b : number density of the ambient gas,
c : UV-photon number luminosity of the central star,
d : FUV-photon number luminosity of the central star,
e : initial Str\"omgren radius, f : dynamical time, 
$R_{\rm st}/C_{\rm II}$, where $C_{\rm II}$ is the sound speed at 
a temperature of $T = 10^4$~K.
\end{table}
 
We calculate the time evolution of the expanding H~II regions
in both the CNM and WNM. For the ambient CNM and WNM, we adopt the
typical number densities of $n_{\rm H,0} = 10~{\rm cm}^{-3}$ and
$n_{\rm H,0} = 0.1~{\rm cm}^{-3}$ respectively.  
The equilibrium temperature and ionization fraction are
$106~{\rm K}$ and $5.9 \times 10^{-4}$ in the CNM, and 
$7900~{\rm K}$ and $2.9 \times 10^{-2}$ in the WNM.
The time evolution is calculated until 10~Myr in all models.
In order to focus on the role of 
stellar UV and FUV radiation, we do not include other kinetic effects 
produced by wind-driven bubbles and supernova explosions.
We briefly discuss these other dynamical processes in \S~\ref{sec:disc}.

\section{Results of the Numerical Calculations}
\label{sec:nres}

\subsection{Expansion in Cold Neutral Medium}
\label{ssec:ex_cnm}

\subsubsection{Fiducial Model : Expansion around $41~M_\odot$ Star}
\label{ssec:fid}

\begin{figure}[tb]
\begin{center}
\includegraphics[width=0.6\hsize]{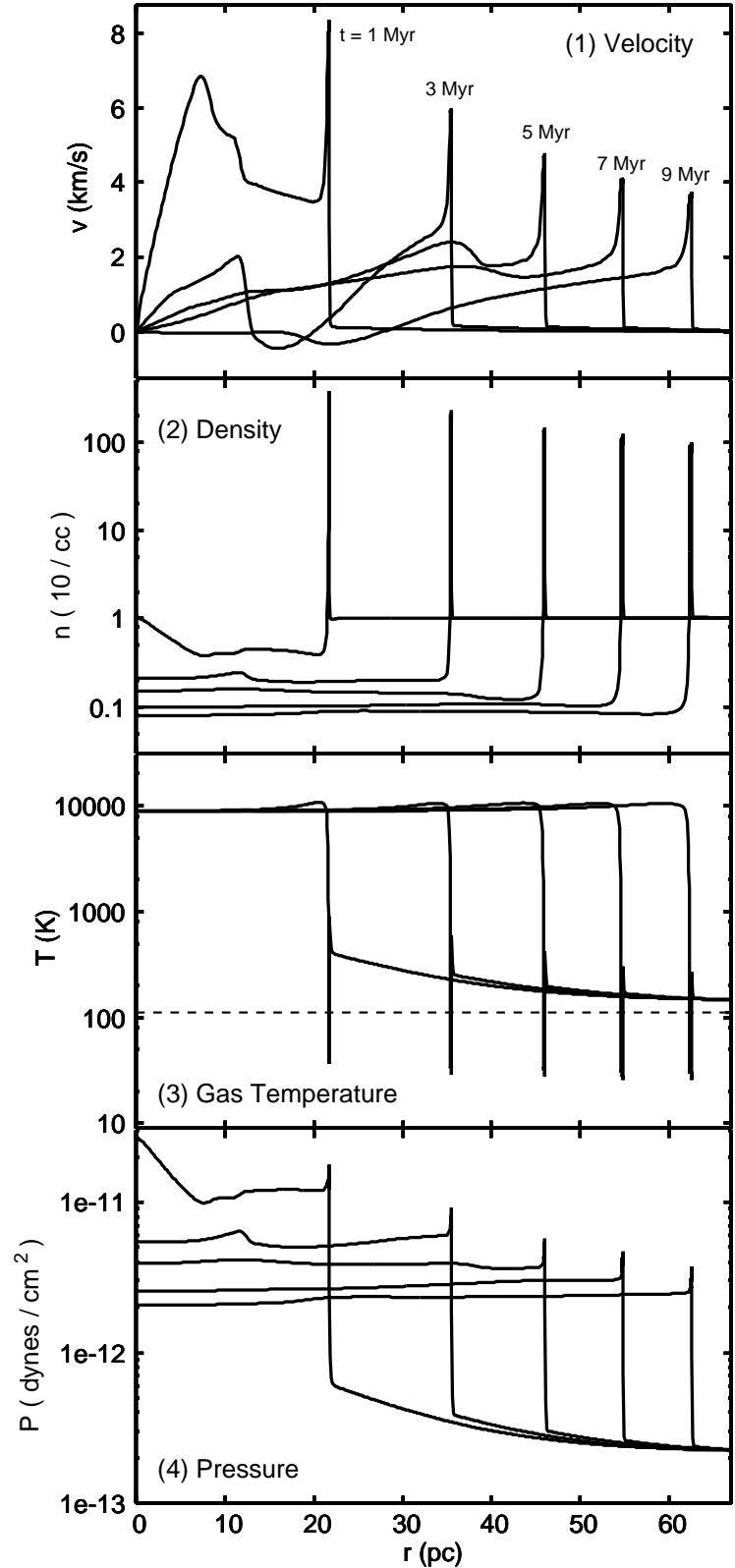}
\caption{Snapshots of the gas-dynamical evolution of model
CNM-S41. In each panel, five snapshots
represent the profiles at $t = 1$, 3, 5, 7 and 9~Myr respectively.
The broken line in panel (3) represents the initial temperature 
of the ambient neutral medium.
 }
\label{fig:hev_s3_cnm}
\end{center}
\end{figure}

First, we consider the time evolution around a massive star
of $41~M_\odot$ in the ambient CNM (model CNM-S41) as the fiducial model.
Figure \ref{fig:hev_s3_cnm} shows the hydrodynamical evolution of 
the H~II region and shell in this model. The basic time evolution is 
the same as that of the H~II region in the ambient molecular material 
(Papers I and II). When the IF passes the initial Str\"omgren 
radius ($\sim 10$~pc), the SF emerges in front of the IF.
Since the gas temperature of the ambient CNM is only $\sim 100$~K,
the H~II region ($T \sim 10^4$~K) has sufficient excess
pressure to drive the SF. The SF sweeps up the ambient 
CNM, and a dense shell is formed just around the 
H~II region. The time evolution of the expansion is well approximated by,
\begin{equation}
R_{\rm IF}(t) = R_{\rm st} 
                \left( 
                         1 + \frac{7}{4} \sqrt{\frac43}
                             \frac{t}{t_{\rm dyn}}
                \right)^{4/7} .
\label{eq:rif}
\end{equation}
The characteristic length scale, $R_{\rm st}$ is the initial 
Str\"omgren radius,
\begin{equation}
R_{\rm st} = 13.9~{\rm pc}~
             \left( \frac{S_{\rm UV}}{10^{49}~{\rm s^{-1}}}
             \right)^{1/3}
             \left( 
              \frac{n_0}{10~{\rm cm}^{-3}} 
             \right)^{-2/3}, 
\label{eq:rst}
\end{equation}
and timescale, $t_{\rm dyn}$ is the dynamical time,
\begin{equation}
t_{\rm dyn} = R_{\rm st}/C_{\rm II} 
            = 1.22~{\rm Myr}~
              \left( \frac{S_{\rm UV}}{10^{49}~{\rm s^{-1}}}
             \right)^{1/3}
             \left( 
              \frac{n_0}{10~{\rm cm}^{-3}} 
             \right)^{-2/3}  ,
\label{eq:tdyn}
\end{equation}
where $S_{\rm UV}$ is the UV-photon number luminosity 
of the central star, $C_{\rm II}$ is the sound speed in
the H~II region, and $n_0$ is the ambient number density.
The gas density in the shell is about 100 times as dense as the 
ambient CNM. The gas temperature in the shell is only $\sim 30$~K, 
which is much lower than the ambient temperature.
The ambient neutral medium just outside the H~II region is slightly
warmer than that in the outermost region. This is due to photoelectric 
heating by the intense FUV radiation from the central star. 
For example, the FUV flux just outside the SF is $G_{\rm FUV} \sim 10$ 
at $t \sim 1$~Myr. Because of the geometrical dilution, the FUV field 
outside the SF gradually decreases as the HII region expands. 

\begin{figure}[tb]
\begin{center}
\includegraphics[width=0.85\hsize]{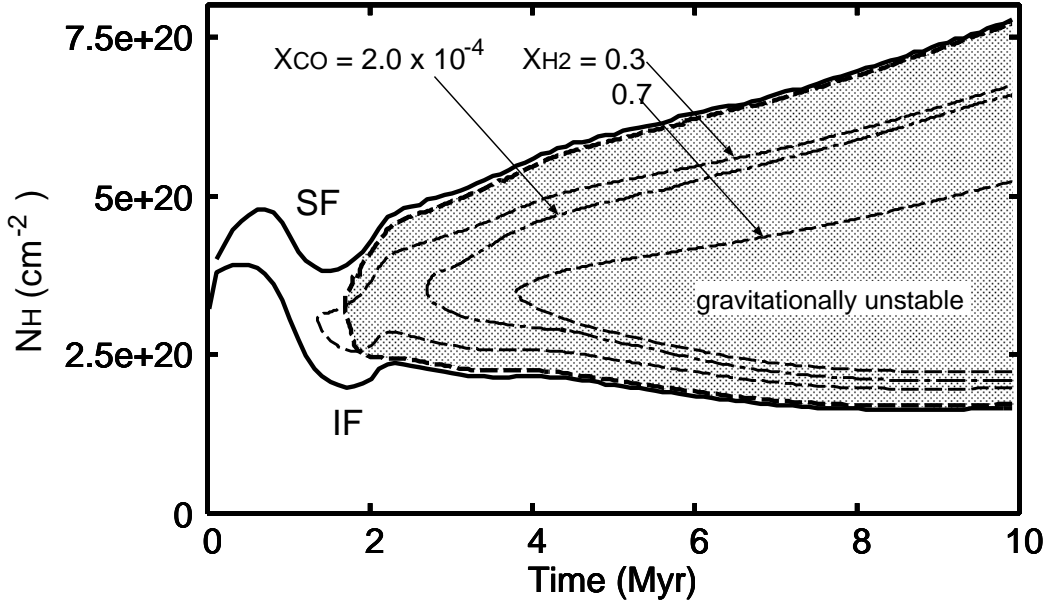}
\caption{Time evolution of the column density of each region
in model CNM-S41. The vertical axis represents the column density 
of the hydrogen nucleon from the central star, and horizontal 
axis represents the elapsed time of the calculation. 
Two thick solid lines correspond to the positions of the SF and IF, 
and the region enclosed between these lines corresponds to the shell. 
Although the radius of the IF monotonically increases, the
column density of the H~II region decreases owing to the decrease
of the density. 
Thin contour curves represent the chemical abundance in the shell;
positions where $X_{\rm H_2} \equiv 2 n_{\rm H_2}/n_{\rm H_{nuc}} 
= 0.3$ and 0.7 (broken), and 
$X_{\rm CO} \equiv n_{\rm CO}/n_{\rm C_{nuc}} = 2.0 \times 10^{-4}$ 
(dot-solid). The shaded region denotes the gravitationally unstable 
region, where $t > (G \rho)^{-0.5}$.}
\label{fig:ctur_s3}
\end{center}
\end{figure}

Figure \ref{fig:ctur_s3} shows the time evolution of the chemical
structure of the shell. The hydrogen molecules gradually accumulate 
in the shell, owing to the rapid reformation on the surface of 
dust grains. About 45\% of hydrogen atoms within the shell
exist as H$_2$ molecules at $t \sim 4.5$~Myr, which is about
the lifetime of the adopted main-sequence star. On the other hand, 
carbon monoxide molecules hardly form in the shell. 
The reason for this difference is that in this model, the column 
density of the shell is only 
$N_{\rm sh} \sim 2.5-5 \times 10^{20}~{\rm cm}^{-2}$, which
corresponds to A$_{\rm V} \sim 0.125-0.25$.
This means that the incident FUV radiation is not significantly
attenuated by the dust absorption in the shell.
Since the dust absorption does not work, the reformed molecules 
are protected against FUV photons only by the self-shielding effect. 
In the equilibrium photodissociation region (PDR) where the
self-shielding is dominant over the dust absorption, the column density 
of the hydrogen nucleon between the IF and H$_2$ dissociation front (DF) is,
\begin{equation}
N_{\rm DF} \sim 5 \times 10^{20}~{\rm cm}^{-2} 
                \left(
                      \frac{\chi}{0.04}
                \right)^{4/3} , 
\label{eq:ndf}
\end{equation}
where $\chi \equiv G_{\rm FUV}/n$ \citep{HT99}.
For the accumulation of H$_2$ molecules in the shell,
$N_{\rm DF}$ must be smaller than the column density 
of the shell, which corresponds to 
$\chi \leq 2-4 \times 10^{-2}$. In our calculation, $\chi$ is 
always only a few $\times 10^{-3}$ at the IF (see also Figure \ref{fig:ng}), 
and the reformed H$_2$ molecules are reliably self-shielded in the shell.
In addition to the sufficient self-shielding in the equilibrium state, 
the H$_2$ reformation timescale is sufficiently short in the shell.
The reformation time is given by
\begin{equation}
t_{\rm form} \sim 100~{\rm Myr}~
                  \left( \frac{n}{10~{\rm cm}^{-3}} \right)^{-1}
\label{eq:rfm}
\end{equation}
\citep[e.g.,][]{HWS71}, and $t_{\rm form} \sim 1$~Myr
in the shell. Therefore, equilibrium H$_2$ abundance is achieved 
during the expansion. On the other hand, the self-shielding of CO 
molecules is inefficient owing to the small abundance of carbon and
oxygen atoms. Even if H$_2$ molecules can accumulate in the shell, 
CO molecules cannot form without sufficient $N_{\rm sh}$ for shielding 
FUV photons by the dust absorption \citep[][Paper II]{Bg04}.
The shielding by H$_2$ molecules is an effect secondary to the
dust shielding. These features do not change in the H~I clouds 
with $n_{\rm H,0} \sim$ a few $\times 10~{\rm cm}^{-3}$, but change 
in the molecular clouds, where the ambient number density is high, 
$n_{\rm H,0} \geq 100~{\rm cm}^{-3}$.
This is well understood with the $n$-dependence of the column 
density of the shell, $N_{\rm sh}$ and the incident FUV flux 
at the IF, $F_{\rm FUV,i}$ at a given $t/t_{\rm dyn}$,
\begin{equation}
N_{\rm sh} \propto n R_{\rm st} \propto n^{1/3} ,
\label{eq:nsh}
\end{equation}
\begin{equation}
F_{\rm FUV,i} \propto S_{\rm FUV}/R_{\rm st}^2 
              \propto n^{4/3} .
\label{eq:ffuvi}
\end{equation}
As equation (\ref{eq:nsh}) shows, 
the column density of the shell rises with the higher 
ambient number density. Although equation (\ref{eq:ffuvi}) shows 
that the incident FUV flux also increases with increasing $n$, 
the intense FUV radiation can be shielded.
Since the dust attenuation works as $\exp(- \tau_d)$, the dust
shielding effect quickly becomes significant once the dust opacity 
of the shell, $\tau_d~(\propto N_{\rm sh})$ exceeds unity.
When the H~II region sufficiently expands in the molecular cloud, 
therefore, the FUV radiation can be blocked by the dust absorption, 
and both H$_2$ and CO molecules can accumulate in the shell 
(see Papers I and II). 

\begin{figure}[tb]
\begin{center}
\includegraphics[width=0.8\hsize]{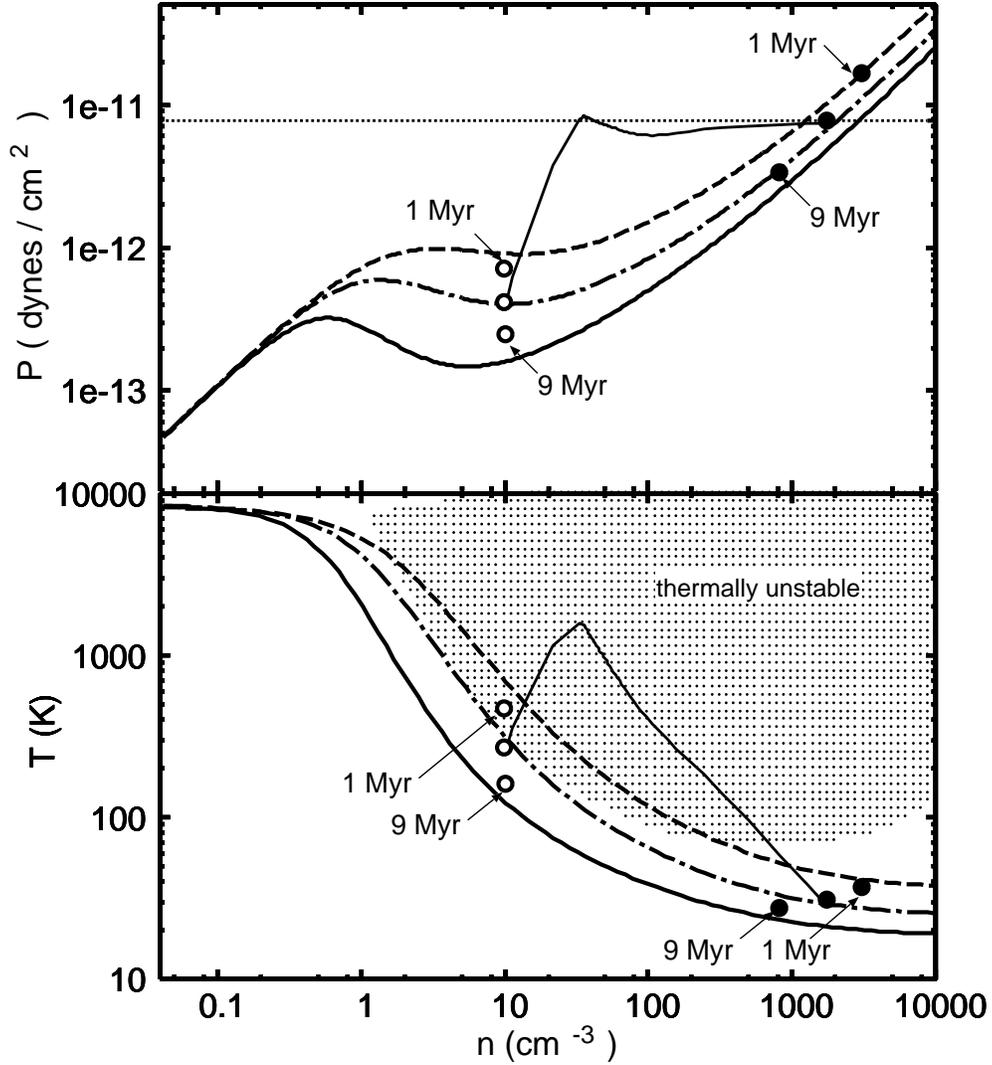}
\caption{The radiative equilibrium states shown on the 
n-P (upper panel) and n-T (lower panel) 
planes. Solid, dot-solid, and broken lines represent the radiative
equilibrium curves for different FUV fields, $G_{\rm FUV}=1$, 3.3, and 10
respectively. In each panel, ``$\circ$''(``$\bullet$'') signs indicate
the calculated states in the pre-shock (post-shock) region 
at $t = 1$, 3 and 9 Myr in model CNM-S41.
In the upper panel, the dotted line indicates $P_{\rm ram}$
at $t = 3$~Myr calculated by equation (\ref{eq:pramt}).
The evolutionary track across the SF at $t \sim 3$~Myr is shown
by thin solid line in each panel. 
In the lower panel, the shaded region denotes where the
isobarically contracting and cooling layer is expected to be
thermally unstable by the linear stability analysis (see text). }
\label{fig:eq_s3}
\end{center}
\end{figure}

In our calculation, the swept-up CNM ($T \sim 100$~K) is converted 
to the cold ($T \sim 30$~K) partially atomic and molecular gas 
behind the SF. The formation mechanism of such a dense and cold shell 
is not understood within the framework of the isothermal SF.
Furthermore, Figure \ref{fig:hev_s3_cnm} shows that the shell density
and temperature gradually decrease as the H~II region expands.
We can understand all these features with the equilibrium $n$-$P$ 
and $n$-$T$ curves of the neutral medium.  
In Figure \ref{fig:eq_s3}, ``$\circ$'' signs represent the physical 
states in the pre-shock region at $t=1,3$, and 9~Myr in our numerical
simulation. This figure shows that ``$\circ$'' sign vertically 
goes down with increasing time $t$, which means that the pre-shock
temperature and pressure decrease almost isochorically.
This is because the FUV radiation field just outside
the shell gradually diminishes owing to the geometrical dilution, so that
the photoelectric heating becomes inefficient.
Since the heating timescale is much shorter than the dynamical
timescale, $t_{\rm dyn} \sim 1$~Myr, the thermal equilibrium
is achieved at each time step.
In the same figure, ``$\bullet$'' signs represent physical states
in the post-shock region at each snapshot.
The upper panel shows that the gas pressure increases
by about ten times across the SF.
We can understand this high post-shock pressure by recalling
the derivation of equation (\ref{eq:rif}). 
Equation (\ref{eq:rif}) is derived from the equation of motion of 
the shell, 
\begin{equation}
M_{\rm sh} \ddot{R}_{\rm IF} = 4 \pi R_{\rm IF}^2 
(P_{\rm II} - P_{\rm ram}) 
\label{eq:eos}
\end{equation} 
(Paper II).
In this equation, $M_{\rm sh}$ is the mass of the swept-up shell,
\begin{equation}
M_{\rm sh} = \frac{4 \pi}3 \rho_0 R_{\rm IF}^3 ,
\end{equation}
$P_{\rm II}$ is the thermal pressure in the H~II region,
\begin{equation}
P_{\rm II} = \rho_0 C_{\rm II}^2 
             \left( 
              \frac{R_{\rm st}}{R_{\rm IF}}
             \right)^{3/2} ,
\label{eq:pii}
\end{equation}
$P_{\rm ram}$ is the ram pressure,
\begin{equation}
P_{\rm ram} = \rho_0 \dot{R}_{\rm IF}^2 ,
\label{eq:pram}
\end{equation} 
and $\rho_0$ is the ambient mass density.
As equation (\ref{eq:eos}) shows, the motion of the shell is
controlled by the balance between the thermal and ram pressure from the 
inner and outer edge of the shell. 
Using equations (\ref{eq:rif}), (\ref{eq:pii})
and (\ref{eq:pram}), the ram pressure can be written as,
\begin{equation}
P_{\rm ram} = \frac{4}{3} \rho_0 C_{\rm II}^2
              \left(
               1 + \frac{7}{4} \sqrt{\frac43} \frac{t}{t_{\rm dyn}}
              \right)^{-4/7}
            = \frac43 P_{\rm II} .
\label{eq:pramt}
\end{equation}
The backward ram pressure is always greater than the forward thermal 
pressure by a factor of $4/3$, and the expansion is decelerating.
Since the sound crossing time of the shell is much shorter than 
$t_{\rm dyn}$, the pressure within the shell smoothly connects 
with $P_{\rm ram}$ ($P_{\rm II}$) at the outer (inner) edge of the 
shell. Therefore, the post-shock pressure is almost equal to 
$P_{\rm ram}$ at each time step.
The pressurized neutral medium is not initially in the thermal
equilibrium state, but going to the equilibrium state over the
cooling timescale, $t_{\rm cool}$.
Since $t_{\rm cool} \sim 0.1$~Myr is much shorter than 
$t_{\rm dyn}$, the gas element engulfed in the shell quickly 
reaches the thermal equilibrium state.
Figure \ref{fig:eq_s3} shows that the calculated density and 
temperature in the post-shock region are equal to the equilibrium
values for the given post-shock pressure ($\sim P_{\rm ram}$)
and FUV field at each time step.

While the gas element is going to the new equilibrium
state, the total cooling rate significantly exceeds the 
heating rate. The dominant cooling process for this 
non-equilibrium gas is the line cooling of CII~(157.7$\mu$m).
The frequency-integrated emergent intensity of this line 
emission is about,
\begin{equation}
I_{\rm ne} \sim \frac{1}{8 \pi} \rho_0 v_{\rm sh}^3
= 0.9 \times 10^{-6} 
  \left( \frac{n_{\rm H,0}}{10~{\rm cm}^{-3}}  \right)
  \left( \frac{v_{\rm sh}}{10~{\rm km/s}} \right)^3
  \quad {\rm ergs / cm^2 / sec / str}
\label{eq:ine}
\end{equation}
\citep{HM79}, where $\rho_0$ is the ambient mass density,
and $v_{\rm sh}$ is the shock velocity.
As equation (\ref{eq:ine}) shows, the emission from the 
non-equilibrium cooling layer quickly diminishes as the 
shell decelerates.
Besides the non-equilibrium cooling layer just behind
the SF, most of the gas within the shell is in the
thermal equilibrium state, where the photoelectric
heating is mainly balanced by the CII~(157.7$\mu$m) line cooling. 
Therefore, the CII line emission also comes from the shell itself.
The emergent intensity from the equilibrium shell is about,
\begin{equation}
I_{\rm e} \sim \frac{1}{2 \pi} \overline{\Lambda_{\rm CII}} N_{\rm sh}
= 1.6 \times 10^{-6} 
  \left( \frac{\overline{\Lambda_{\rm CII}}}{10^{-25} {\rm ergs/sec}} 
  \right)
  \left( \frac{N_{\rm sh}}{10^{20}~{\rm cm}^{-2}} \right) 
  \quad {\rm ergs / cm^2 / sec / str},
\label{eq:ie}
\end{equation}
where $\overline{\Lambda_{\rm CII}}$ is the average CII cooling
rate in the shell. As expected from equations (\ref{eq:ine}) and 
(\ref{eq:ie}), the total CII emission is dominated by that from 
the shell in the thermal equilibrium, but for the very early phase
of the expansion.
Although this far-IR CII emission could be detected,
this line, an ordinary feature of the PDR,
is not a good signpost of the molecular cloud formation.
Instead, we seek the predicted gas phase using the HISA 
feature in \S~\ref{sec:obs}.

In our 1-D calculation, the neutral medium is steadily swept up 
into the shell. However, some instabilities actually deform the shell 
and excite its internal structure.
First, the dense gas layer is expected to fragment by the gravitational
instability. The shell-fragmentation occurs first at the length scale 
of the layer thickness, with a growth timescale of  
$\sim (G \rho)^{-1/2}$ \citep[e.g.,][]{EE78,MNH87,NIM98}.
We adopt $t > (G \rho)^{-1/2}$ as the simplified criterion
of the gravitationally unstable layer. 
Figure \ref{fig:ctur_s3} shows that the shell becomes gravitationally
unstable soon after $t \sim 2$~Myr, when H$_2$ molecules begin to 
accumulate in the shell. The typical radius and mass of the fragment 
are about 0.1~pc and $10~M_\odot$ respectively.  
If each fragment subsequently contracts, the molecular reformation
will be promoted by the self-shielding of cores, owing to the increased 
column density. Second, the compressed layer collapsing 
due to the significant radiative cooling just behind the SF, suffers 
from thermal instability \citep{ki00,ki02}.
In Figure \ref{fig:eq_s3}, we plot the evolution track of a fluid element 
across the SF at $t \sim 3$~Myr.  The gas temperature rises up to 
$T \sim 2000$~K at the SF, due to compressional heating, and falls 
to $T \sim 30$~K, due to the significant radiative cooling behind the SF. 
The timescale to settle into the new equilibrium state is about the 
cooling time, $t_{\rm cool} \sim 0.1$~Myr.
The upper panel of Figure \ref{fig:eq_s3} shows that the collapsing layer
evolves almost isobarically.
The linear stability analysis of the isobarically collapsing layer
has been explored by \citet{ki00}, and the shaded region in the 
lower panel of Figure \ref{fig:eq_s3} represents
the expected thermally unstable region with $G_{\rm FUV} = 3.3$.
Note that the gas is cooling and isobarically contracting in the 
unperturbed state in their linear analysis. 
Whether the small perturbation grows or not is analyzed in the 
evolving background state, which differs from the original analysis
by \citet{Fd65}.
Figure \ref{fig:eq_s3} shows that the compressed CNM becomes 
thermally unstable on its way to the new equilibrium state.
The linear analysis shows that the layer fragments 
into tiny cloudlets, whose size is about tens of AUs, over the 
timescale of $10^{3-4}$~yr \citep{ki00}.
Multi-dimensional numerical simulations \citep[e.g.,][]{ki02,ah05}
show that each fragment has some translational motion of a few
km/s. This motion is driven by the pressure of the
surrounding warmer, less dense region. 
Thus, the motion is inevitably subsonic for the surrounding
medium, and does not dissipate over the sound-crossing time.
\citet{ki02} have suggested that this translational cloudlets 
motion is the origin of the observed
broad line width, or ``turbulence'' in the cold clouds.
Our current 1-D calculations cannot explain the the actual evolution 
of the fragmentation of the shell, which will be affected by both the 
thermal and gravitational instabilities. Future multi-dimensional 
simulations of radiative-hydrodynamics will reveal the detailed nature of
the fragmented shell.

\subsubsection{Dependence on Central Stars}
\label{ssec:dep_star}

Second, we study how time evolution changes with different central 
stars in the same ambient CNM. Our numerical results show a clear feature;
with the higher-mass central star, the shell density is higher
and the reformation of H$_2$ molecules becomes more efficient 
in the shell.

\begin{figure}[tb]
\begin{center}
\includegraphics[width=0.6\hsize]{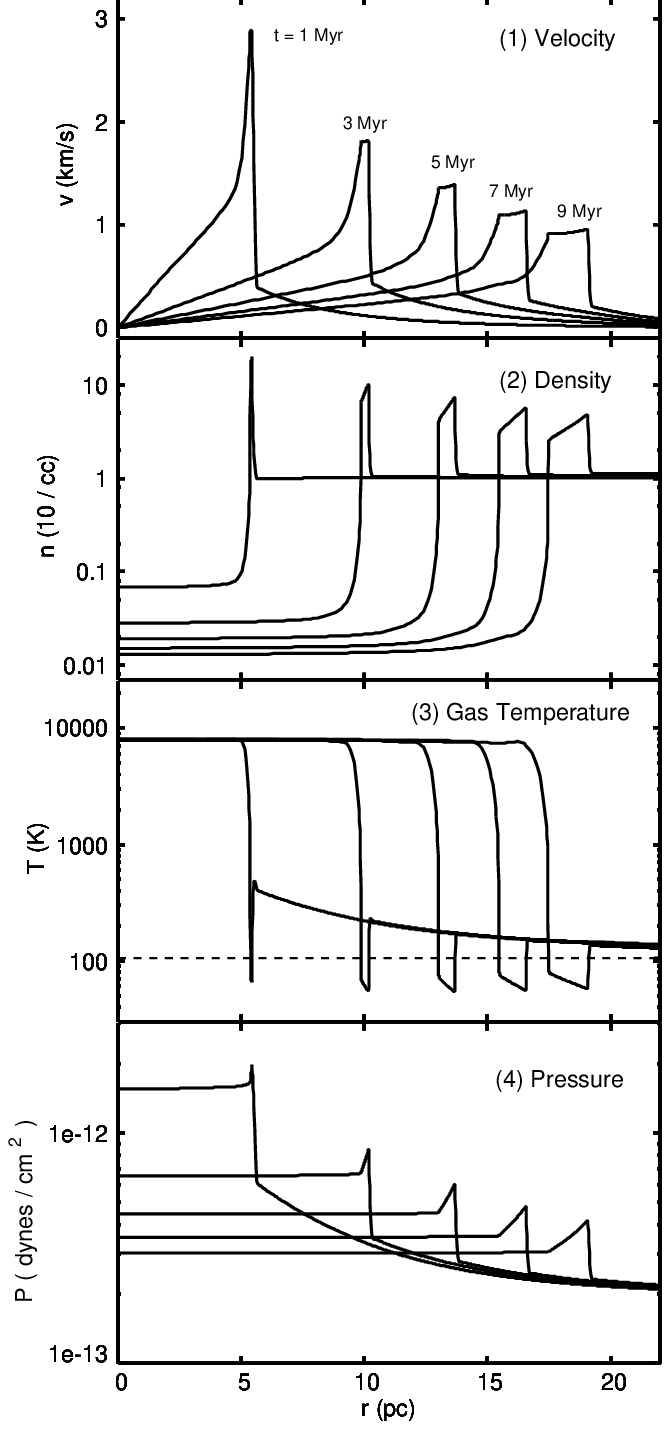}
\caption{Same as Figure \ref{fig:hev_s8_cnm} but for model CNM-S12.
 }
\label{fig:hev_s8_cnm}
\end{center}
\end{figure}

\begin{figure}[tb]
\begin{center}
\includegraphics[width=0.85\hsize]{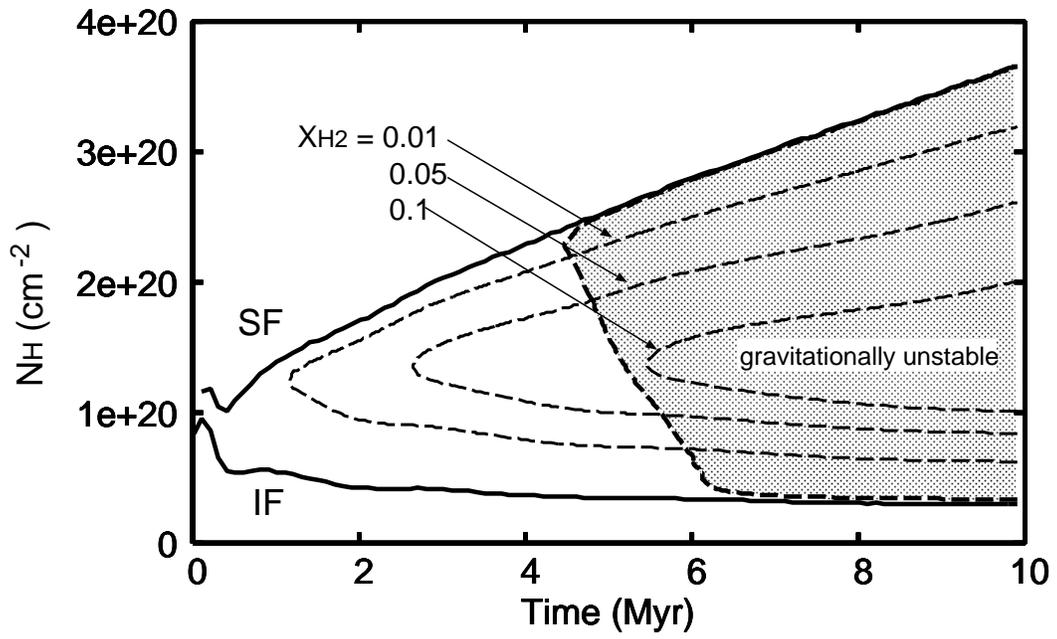}
\caption{Same as Figure \ref{fig:ctur_s3} but for model CNM-S18.
The contours with thin broken curves represent the positions where
$X_{\rm H_2}=0.01$, 0.05 and 0.1 respectively.}
\label{fig:ctur_s6}
\end{center}
\end{figure}

\begin{figure}[tb]
\begin{center}
\includegraphics[width=0.7\hsize]{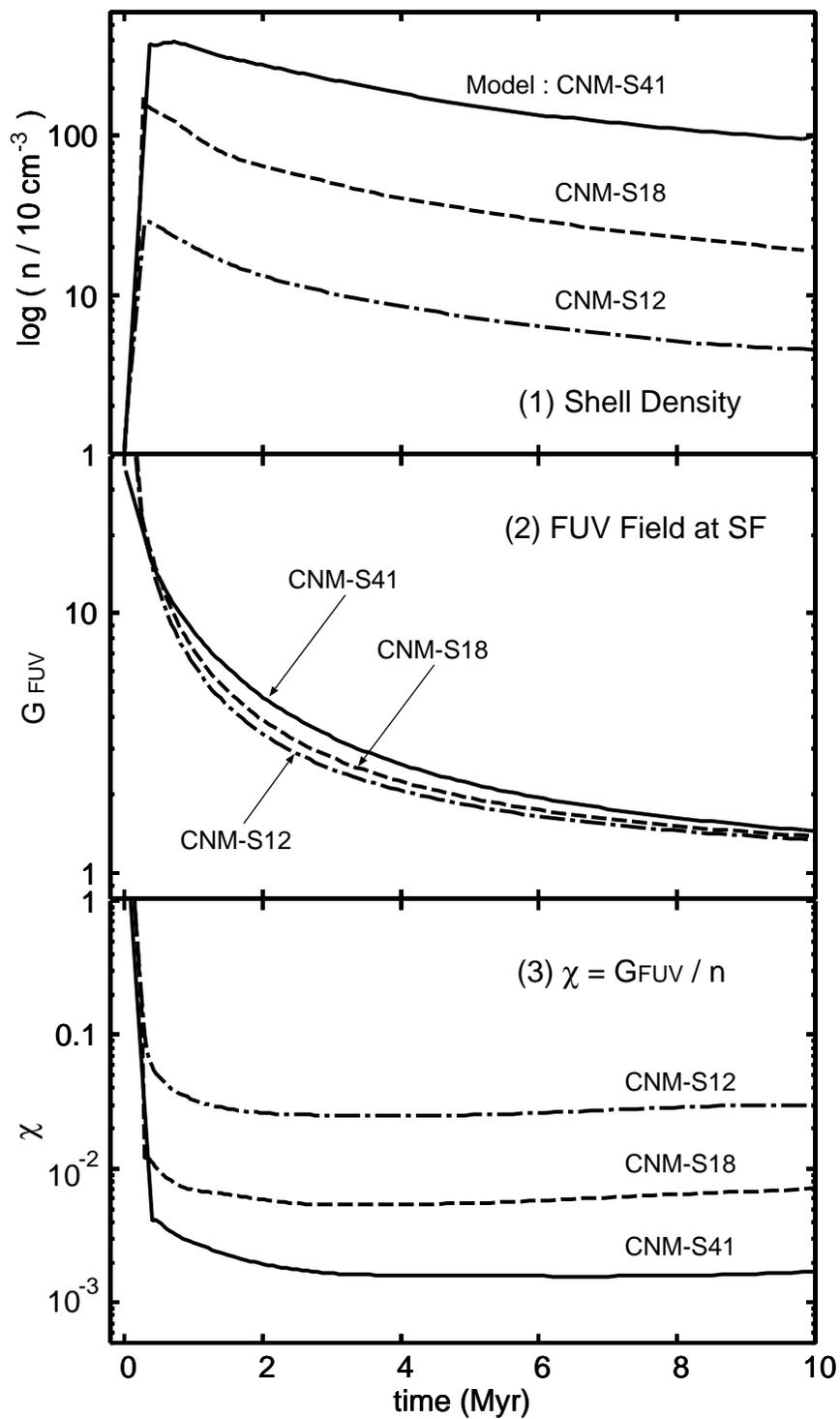}
\caption{
Time evolution of the density of the shell, FUV radiation field
at the SF, and their ratio, $\chi$.
The solid, broken, and dot-solid curves in each panel represent
the evolution of models CNM-S41, S18, and S12 respectively.
}
\label{fig:ng}
\end{center}
\end{figure}

For example, Figure \ref{fig:hev_s8_cnm} presents gas-dynamical 
evolution around a massive star of $M_* = 11.7~M_\odot$ 
(model CNM-S12). The SF emerges and sweeps up the ambient CNM, 
which is the same as model CNM-S41.
The shell density is 10 times as high as the ambient 
CNM density at $t \sim 1$~Myr, but quickly declines afterward. 
The shell density is always lower than that in model CNM-S41 at
the same time, $t$ (c.f. Figure \ref{fig:ng}). 
Despite the different densities of the shells, the temperature
profiles in Figures \ref{fig:hev_s3_cnm} and \ref{fig:hev_s8_cnm}
are similar. This is because the FUV field outside the shell is 
similar among these models (Figure\ref{fig:ng}). 
The FUV flux just outside the SF is $G_{\rm FUV} \sim 10$ at 
$t \sim 1$~Myr, and gradually decreases owing to the geometrical dilution.   
The formation of molecules in the shell hardly occurs in 
model CNM-S12. No CO molecules are formed, and only less than 0.1\% 
of H atoms is converted to H$_2$ molecules in the shell over 10~Myr. 
Since the column density of the shell is 
$N_{\rm sh} < 2.5 \times 10^{20}~{\rm cm}^{-2}$, the dust 
absorption does not protect reformed molecules against FUV photons.
As shown in equation (\ref{eq:ndf}), the self-shielding becomes
less efficient with the larger $\chi = G_{\rm FUV}/n$ in 
the shell. In model CNM-S12, $\chi$ is about ten times 
as large as that in model CNM-S41 at the same time, $t$ 
(also see Figure \ref{fig:ng}). 
Therefore, $N_{\rm DF}$ given by equation (\ref{eq:ndf}) 
exceeds the column density of the shell, and even
the self-shielding of H$_2$ molecules does not work. 
Furthermore, if the self-shielding were efficient, H$_2$ molecules 
would not form in the low-density shell. 
Equation (\ref{eq:rfm}) shows that the reformation timescale 
becomes longer than 10~Myr, when the shell density 
is smaller than $100~{\rm cm}^{-3}$. The reformation timescale is 
so long that no equilibrium H$_2$ abundance is achieved.
We have calculated other models with different exciting stars, 
and found that the formation of H$_2$ molecules is more
efficient with higher-mass stars. Figure \ref{fig:ctur_s6} shows
the time evolution of the chemical structure of the shell 
around the central star of $17.5~M_\odot$ (model CNM-S18). 
In this model, H$_2$ molecules are slightly formed in the shell.
About 6~\% of swept-up hydrogen atoms are converted to H$_2$ 
molecules by the time of 10~Myr. 

We investigate why the efficiency of H$_2$ formation depends 
on the mass, or UV/FUV luminosity of the central star. 
The key physical quantities are the ratio, 
$\chi =  G_{\rm FUV}/n$ and number density, $n$, in the shell. 
The former reflects the efficiency of the self-shielding effect,
which determines the H$_2$ abundance in the equilibrium state, 
and the latter determines the reformation timescale, 
over which the H$_2$ abundance becomes the equilibrium value. 
Figure \ref{fig:ng} shows the time evolution of the number density, 
FUV radiation field, and their ratio, 
$\chi$ in the shell in models CNM-S41, S18, and S12, respectively.
At the same time, $t$, the number density is larger with 
higher-mass stars, and the FUV fields are similar among models.
Thus, their ratio, $\chi$ is smaller with a higher-mass star.
The self-shielding effect is more efficient with a higher-mass 
central star, as equation (\ref{eq:ndf}) shows.
The higher shell density ensures that the equilibrium state is achieved
over a shorter timescale, and that H$_2$ molecules quickly accumulate 
in the shell with a higher-mass star. 
Consequently, the variation in the chemical structure of the shell 
originates in similar and different time evolutions of $G_{\rm FUV}$ 
and $n$ among models shown in Figure \ref{fig:ng}. 

We make use of some analytic relations to interpret these 
characteristic features. First, the FUV field at time 
$t$, $G_{\rm FUV}(t)$ scales with the UV/FUV luminosity of the star,   
\begin{equation}
G_{\rm FUV}(t) \propto \frac{S_{\rm FUV}}{R_{\rm IF}(t)^2}
               \propto
               S_{\rm FUV}^{1/3}
               \left(
                \frac{S_{\rm FUV}}{S_{\rm UV}}
               \right)^{2/3}
               \left( 
                1 + \frac74 \sqrt{\frac43} \frac{t}{t_{\rm dyn}}
               \right)^{-8/7} .
\end{equation}
This shows that $G_{\rm FUV}(t)$ is normalized with the first term,
$S_{\rm FUV}^{1/3} (S_{\rm FUV}/S_{\rm UV})^{2/3}$, 
which is smaller with the more massive star.
On the other hand, the last evolutionary term has an inverse 
dependence, because the dynamical time, $t_{\rm dyn}$ is 
longer with the higher-mass star (see equation (\ref{eq:tdyn})). 
Owing to the cancellation of these opposite dependences, the time
evolution of the FUV field is similar among models with different stars,
as shown in Figure \ref{fig:ng}.   

\begin{figure}[tb]
\begin{center}
\includegraphics[width=0.8\hsize]{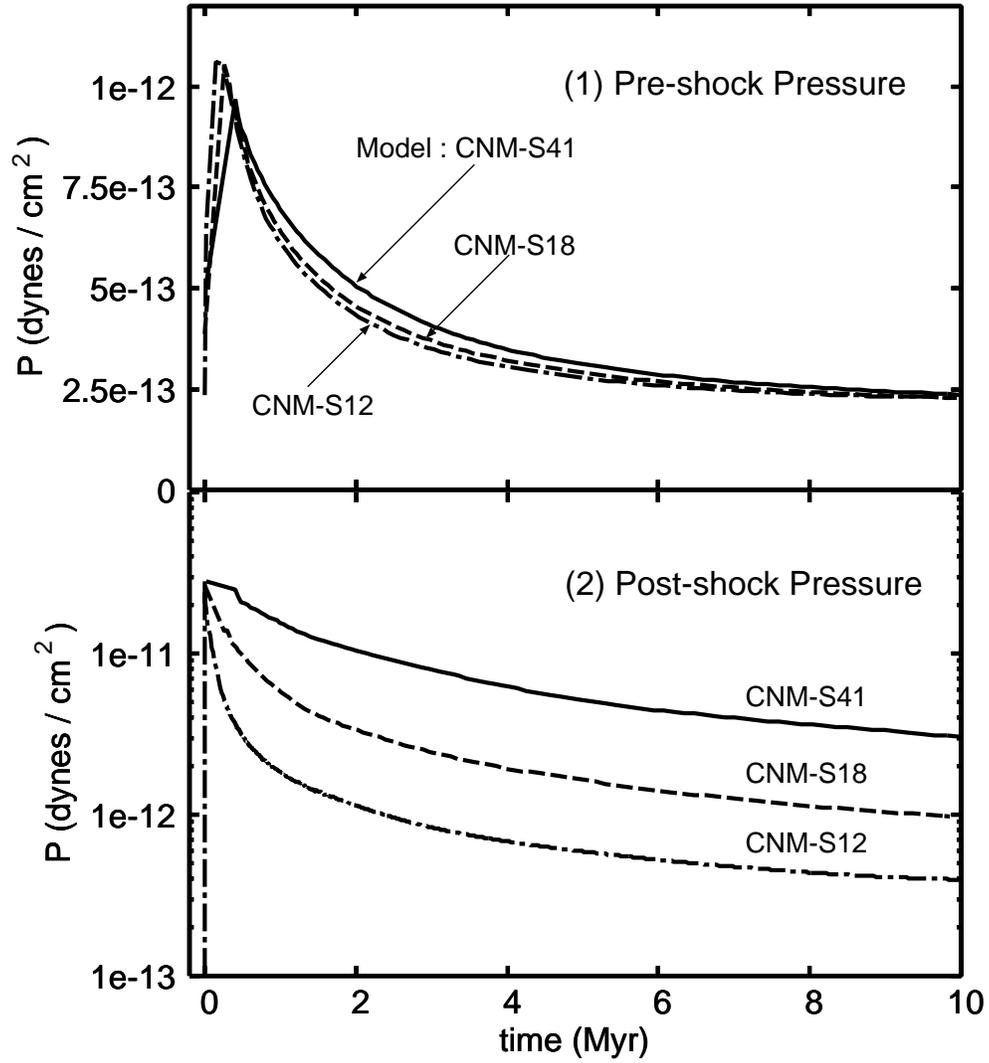}
\caption{
Time evolution of the gas pressure in the pre-shock and post-shock
regions. The solid, broken, and dot-solid lines in each panel represent
the evolution of models CNM-S41, S18, and S12 respectively. 
}
\label{fig:p_acrsf}
\end{center}
\end{figure}

\begin{figure}[tb]
\begin{center}
\includegraphics[width=0.8\hsize]{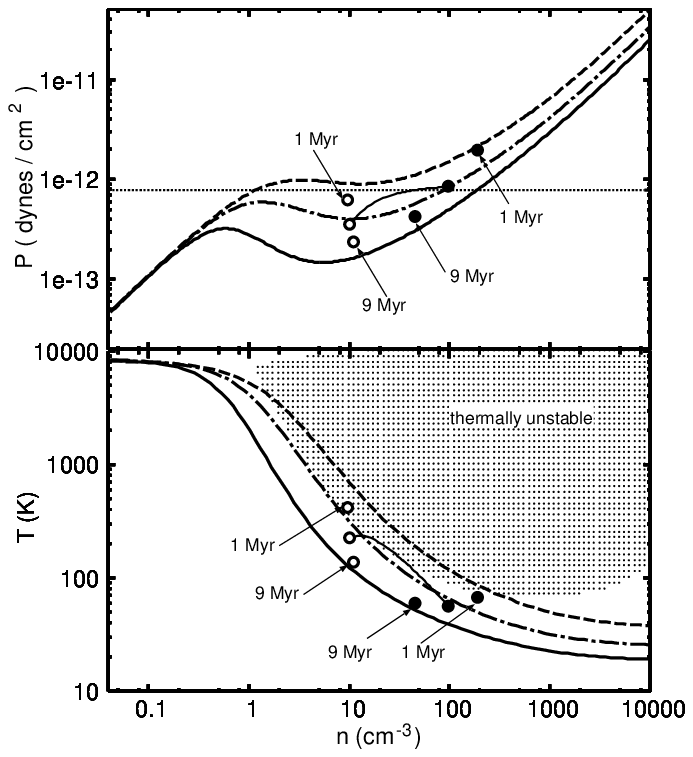}
\caption{Same as Figure \ref{fig:eq_s3} but for model CNM-S12. }
\label{fig:eq_s8}
\end{center}
\end{figure}

Next, let us consider why the shell density is larger with a
higher-mass star. In \S~\ref{ssec:fid}, we have shown that the 
post-shock density is determined by the FUV radiation field and gas 
pressure in the shell at each time step.
Because of the similarity in the time evolution of $G_{\rm FUV}$ 
among models, only gas pressure in the shell dominates the 
post-shock physical state. The time evolution of the ram pressure is 
given by equation (\ref{eq:pramt}), where the dynamical time, 
$t_{\rm dyn}$ is longer with the higher-mass star.
Therefore, thermal pressure in the shell is higher 
with the higher-mass star at the same time, $t$, which
causes a variation in shell density among models.
Figure \ref{fig:p_acrsf} presents the time evolution of 
thermal pressure in both pre-shock and post-shock regions
in models CNM-S41, S18, and S12.
Since the FUV radiation field is similar, the time evolution
of the pre-shock pressure is almost the same among models. 
The post-shock pressure is higher with the higher-mass star,
as predicted by equation (\ref{eq:pramt}). 
The significant enhancement of the pressure behind the SF 
accounts for the high density of the shell.
Figure \ref{fig:eq_s8} presents several post-shock and pre-shock 
physical states in model CNM-S12. 
In this model, the post-shock pressure is just a little higher
than the pre-shock pressure, and the shell density is significantly 
lower than that in model CNM-S41 (c.f. Fig. \ref{fig:eq_s3}).

Finally, we discuss the multi-dimensional structure of the shell 
expected to be produced by various instabilities.
First, gravitational instability grows earlier in denser layers. 
Since the shell density is higher with a higher-mass star,
shell-fragmentation will occur earlier in these higher-mass models.
While the shell becomes unstable at $t \sim 2$~Myr in
model CNM-S41 (Figure \ref{fig:ctur_s3}), it is not until 
$t \sim 5$~Myr that the unstable region appears
in model CNM-S18 (Figure \ref{fig:ctur_s6}). 
No gravitationally unstable region appears in CNM-S12.
Second, the thermal stability of the isobaric cooling layer
depends on the compression level across the SF.
Figure \ref{fig:eq_s8} shows the evolution track
of the layer at $t \sim 3$~Myr in model CNM-S12.
The pressure rises by a factor of two or three across the SF, and
the evolution track only grazes the thermally unstable region
on the $n$-$T$ plane. In summary, our calculations show that 
H$_2$ formation is more efficient in a model with a higher-mass 
star, and that the effects of gravitational/thermal instability 
are more significant in such higher-mass models.

\subsection{Expansion in Warm Neutral Medium}
\label{ssec:ex_wnm}

\begin{figure}[tb]
\begin{center}
\includegraphics[width=0.6\hsize]{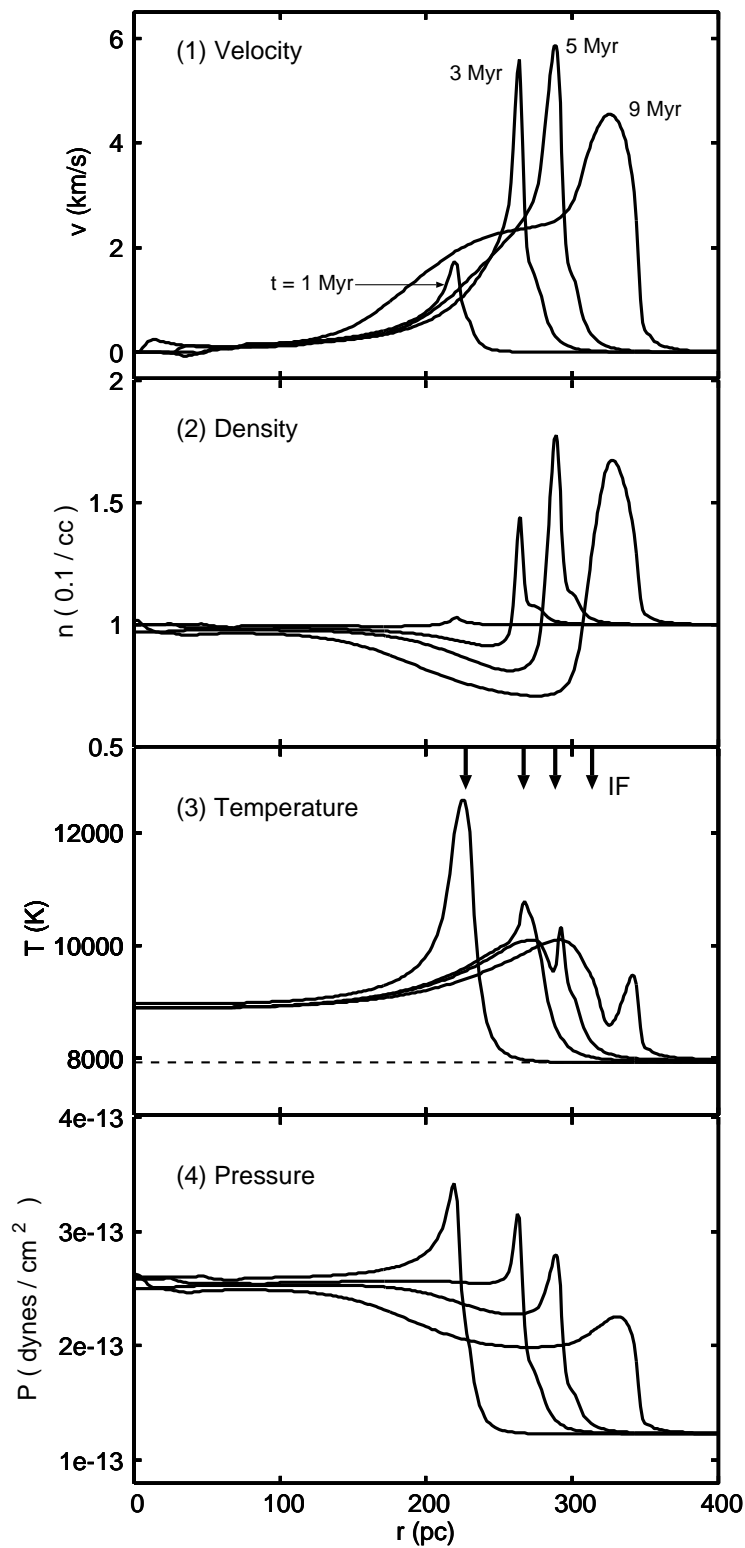}
\caption{Snapshots of the gas-dynamical evolution in model WNM-S41.
In each panel, three snapshots represent the profiles at 
$t = 1$, 3, 5, and 9~Myr respectively.
The broken line and arrows in panel (3) represent the initial equilibrium
temperature of the ambient warm neutral medium, and the position of the 
IF at each time step.}
\label{fig:hev_s3_wnm}
\end{center}
\end{figure}

Once the stellar UV radiation escapes from the molecular clouds
and H~I clouds (CNM), the H~II region expands to the WNM.
Figure \ref{fig:hev_s3_wnm} shows the hydrodynamical evolution
of the H~II region around a $41~M_\odot$ star in the ambient
WNM (model WNM-S41).
Because of the low ambient number density, the H~II region expands 
into a very large region of more than $100$~pc.
As Figure \ref{fig:hev_s3_wnm} shows, however, the expanding H~II 
region hardly affects the hydrodynamics. 
The gas density slightly increases by a factor of two in front of the IF.  
This is partly because of the large initial Str\"omgren
radius. In model WNM-S41, the initial Str\"omgren radius is, 
$R_{\rm st} \sim 250$~pc, and the H~II region remains in its
early formation phase at $t \lesssim 2$~Myr.
Furthermore, after the H~II region reaches the initial Str\"omgren
radius, no dense shell is formed around the H~II region. 
The ambient gas temperature is originally as high as $T \sim 8000$~K, and 
the pressure excess of the H~II region over the ambient WNM 
is small. No SF with a high Mach number ever emerges in front of the IF.
The ambient WNM is only converted to the ionized medium
with a similar temperature and density, which is the WIM.
These basic features of the time evolution do not change with 
different central stars. Although \citet{HP99} and \citet{ki00} 
have shown that the highly pressurized WNM can be converted
into the CNM, this is not the case with the expanding HII regions. 
The bubble driven by stellar-wind or supernova explosion
can provide the much higher pressure even in the WNM, and triggers
the phase transition from WNM to CNM (also see \S~\ref{ssec:bubble}).

\section{Hunt of the Dark Molecular Hydrogen Clouds
         in the Observational Data}
\label{sec:obs}

\subsection{Shell-like HI Self-absorption around Giant H~II Regions}

When the H~II region expands in the CNM, the SF emerges in front
of the IF and the ambient CNM is compressed in the shell.
We have shown that  H$_2$ molecules can form if the central 
star is more massive than $18~M_\odot$, but CO molecules seldom form 
in the shell.  
The calculated temperature and density in the shell are
$T \sim$ a few $\times$ 10~K and $n \sim 100 - 1000~{\rm cm}^{-3}$.
This is just the gas phase intermediate between the ambient CNM and 
molecular clouds observed with the CO emission.
However, it is generally difficult to detect such an intermediate gas phase.
The gas temperature is too low to thermally excite the 
rovibrational transition of H$_2$ molecules. 
The FUV radiation field in the shell is $G_{\rm FUV} \sim 10$ at most, 
and it is hard to detect the UV fluorescent H$_2$ emission from
the distant giant H~II regions.
Instead of hunting for signs of the cold H$_2$ molecules,
we probe the predicted ``dark'' hydrogen with the H~I
self-absorption (HISA).
The HISA is the H~I 21-cm line-absorption against the background
emission from the warmer neutral medium. 
Statistical studies of the CNM usually analyze the distribution of the
21-cm absorption features against the strong radio continuum 
sources, such as compact H~II regions and supernova remnants 
\citep[e.g.,][]{Dc03, HT03a, HT03b}.
On the other hand, the brightness temperature of the background 
H~I 21-cm emission is only $T_{\rm b} \sim 100$~K, and HISA 
features automatically trace the colder tail of the CNM.   

\begin{figure}
  \begin{center}
\includegraphics[width=1.0\hsize]{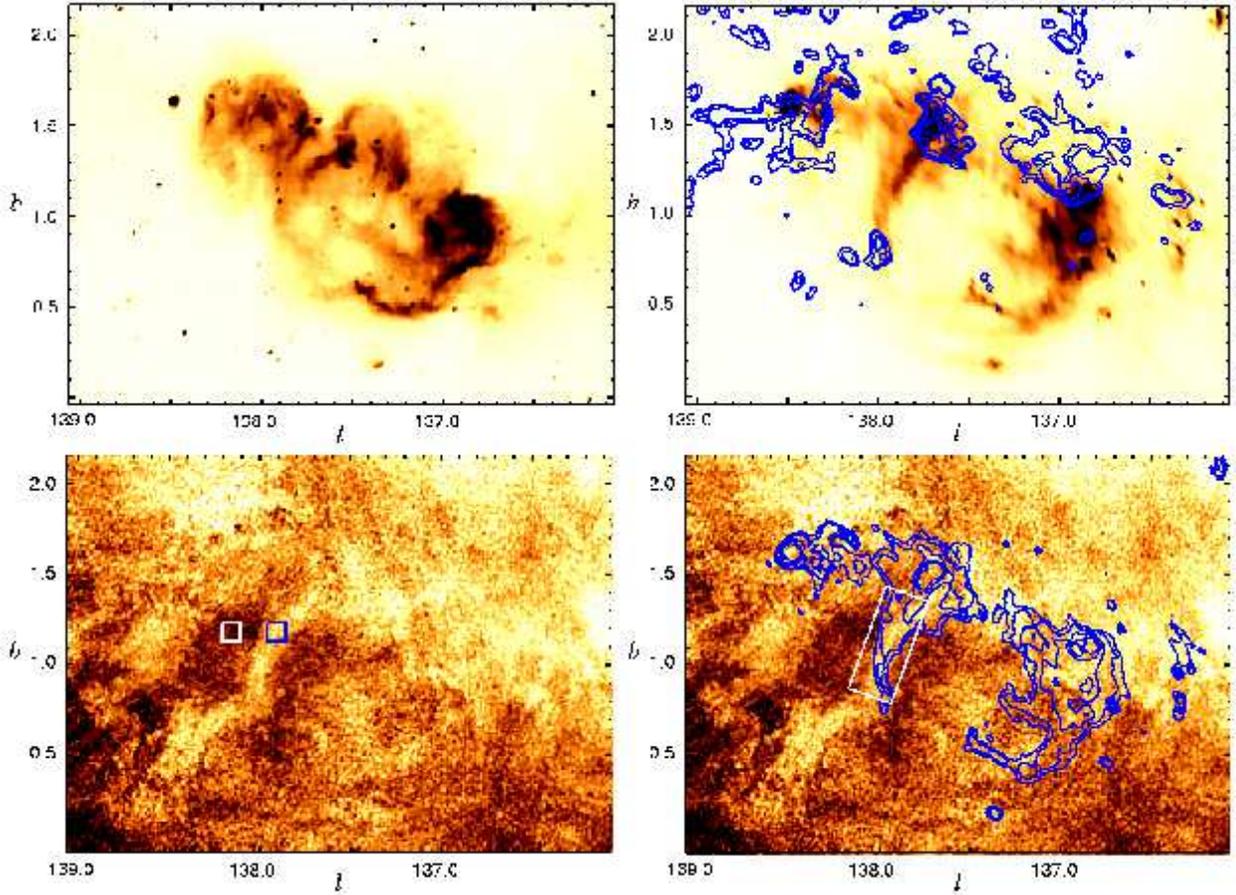}
\caption{
Multi-wavelength view of the W5 region using CGPS data.
{\it Upper left} : image of the 21-cm continuum emission (H~II gas).
The brightness temperature is linearly scaled from 5~K (bright)
to 12.5~K (dark).    
{\it Upper right} : image of the 60-$\mu$m dust emission and 
contours of $^{12}$CO(1-0) line emission at 
$v_{\rm LSR} = -39.8$~km/s. The intensity of the 60-$\mu$m dust 
emission is linearly scaled from 10~MJy/str (bright) to 200~MJy/str (dark).
The contour levels of CO emission are $T_{\rm b} = 1, 2, 5$, and 10~K.
{\it Lower left} : channel map of the H~I 21-cm emission at
$v_{\rm LSR} = -39.8$~km/s. The brightness temperature is linearly 
scaled from 45~K (bright) to 110~K (dark).
The blue (white) $6' \times 6'$ square denotes the on-region 
(off-region) used for the analysis.
{\it Lower right} : channel map of the H~I 21-cm emission at 
$v_{\rm LSR} = -39.8$~km/s and contours of 
60-$\mu$m dust emission. The contour levels are 
$I_{\rm 60 \mu m} = 75, 92.2, 143.8, 230$ and 350~MJy/str.
The region enclosed by the white line denotes where the
HISA feature shows the good spatial correlation with the dust shell.
}
    \label{fig:w5}
  \end{center}
\end{figure}

We have sought the HISA features associated with the giant H~II
regions using recent observational data by the Canadian Galactic
Plane Survey \citep[CPGS; see][]{Tl03}.
The CGPS data provides us with the high-resolution ($\sim 1'$)
maps of HII, HI, CO, and dust in the Galactic plane.
We use the 21-cm continuum image and velocity cube of the 21-cm 
emission by the Dominion Radio Astrophysical Observatory 
(DRAO) survey \citep{HT00}, $^{12}$CO(1-0) data cubes by
the Five College Radio Astronomy Observatory (FCRAO) outer galactic
survey \citep{Hy98}, and reprocessed {\it IRAS} images at 
60 and $100~\mu$m bands \citep{Cao97}. 
We have examined the star-formation complex W3-W4-W5 along the
Perseus arm, which includes the giant H~II regions, W3, W4 and W5.
Adopting a distance of $1.95$~kpc to the Perseus arm \citep{Xu06}, 
the angle of $1\arcdeg$ corresponds to 32.2~pc.  
Figure \ref{fig:w5} shows the multi-wavelength view
of the W5 region. The upper left panel is the 21~cm continuum map, 
which shows distribution of ionized hydrogen.
The upper-right panel presents a $60~\mu$m image
of the W5 region, and we can see the clear shell-like
structure around the ionized gas. Other {\it IRAS}-band images 
also show similar shell-like structures \citep{KM03}.
The lower left panel of Figure \ref{fig:w5} shows a channel map of
21-cm emission at $v_{\rm LSR} = - 39.8$~km/s.
We have also found a narrow shell-like feature in this map,
which shows strong spatial correlation with the dust shell,
as indicated by the white rectangle in the lower right panel.
The brightness temperature along the shell-like feature is 
$T_{\rm b} \sim 70$~K, which is lower than that on both sides of the 
shell, $T_{\rm b} \sim 100$~K.
Since it is very unlikely that the 21-cm emission from the diffuse WNM 
would fluctuate over such a small scale, and that such fluctuation 
would overlap with the dust shell by accident, we consider this 
shell-like feature as being HISA by the colder neutral medium.
Furthermore, these dust shell and shell-like HISA feature should
originate in the same structure just around the W5 H~II region.
We also show the contours of the $^{12}$CO(1-0) channel map at the same 
velocity in the upper right panel, but the CO emission does not trace the
shell-like structures with the current sensitivity of the FCRAO survey. 

\begin{figure}
  \begin{center}
\includegraphics[width=1.0\hsize]{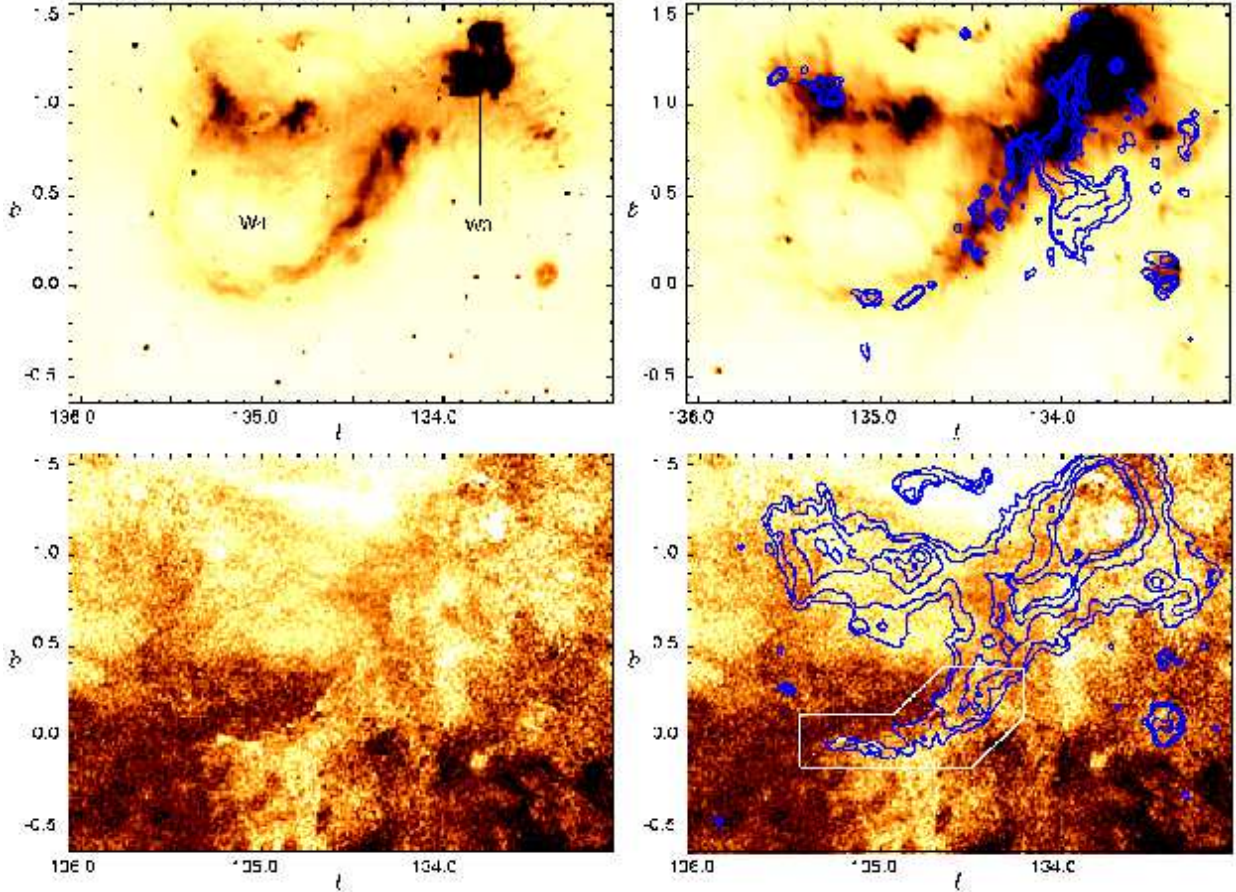}
\caption{Multi-wavelength view of the W3-W4 region using CGPS data. 
The distribution of panels and the scaling of images and contours 
are almost the same as those in Figure \ref{fig:w5}. 
Some differences are as follows:
{\it Upper left} :     The brightness temperature is linearly scaled 
from 4.9~K (bright) to 20.0~K (dark). 
{\it Lower right} : The contour levels of the 60-$\mu$m intensity
of the dust emission are $I_{\rm 60 \mu m} = 65, 82.8, 136.3, 225.3$, 
and 350~MJy/str. The channel maps of $^{12}$CO(1-0) and H~I 21~cm 
emission are at $v_{\rm LSR} = -47.3$~km/s.
The region enclosed by the white line denotes where the
HISA feature shows the good spatial correlation with the dust shell.
}
    \label{fig:w4}
  \end{center}
\end{figure}

We have also found some similar features in the giant H~II region, W4.
Figure \ref{fig:w4} shows a multi-wavelength view of the W3-W4 region. 
The upper left panel shows the distribution of the ionized hydrogen.
The W4 H~II region has an ionized shell-like structure. The 
W3 H~II region is denser and more compact than W4. 
The upper right panel presents the $60~\mu$m image superposed on the
contours of the $^{12}$CO(1-0) channel map at $v_{\rm LSR} = -47.3$~km/s.
The distributions of both the dust and CO emission are two-fold.
In $l \gtrsim 134.3\arcdeg$, the $60~\mu$m emission shows a 
shell-like structure just outside the ionized shell.
The CO molecules are also along the dust shell,
but discretely distributed as small clouds. 
In $\l \lesssim 134.3\arcdeg$, on the other hand,
the broad and strong $60~\mu$m emission comes from the warmer
dust around the W3 H~II region. 
The CO emission also broadly extends in $\l \lesssim 134.3\arcdeg$.
This is the eastern edge of the W3 giant molecular cloud (GMC), which 
actually extends over $l < 133\arcdeg$.
The lower-left panel of Figure \ref{fig:w4} shows the 21-cm channel
map at $v_{\rm LSR} = -47.3$~km/s. The HISA feature has also been found,
and overlaps with the dust shell (lower-right panel).
\citet{CHS00} have shown that the small molecular clouds are widely
distributed in the W3-W4-W5 cloud complex, and have speculated that
these are remnants of pre-existing molecular clouds
dispersed by the negative feedback from massive stars. 
We suggest that a number of the small clouds form from
the compressed ambient neutral medium.

\subsection{Quantitative Analysis of the Data}

\begin{figure}[tb]
\begin{center}
\includegraphics[width=0.7\hsize]{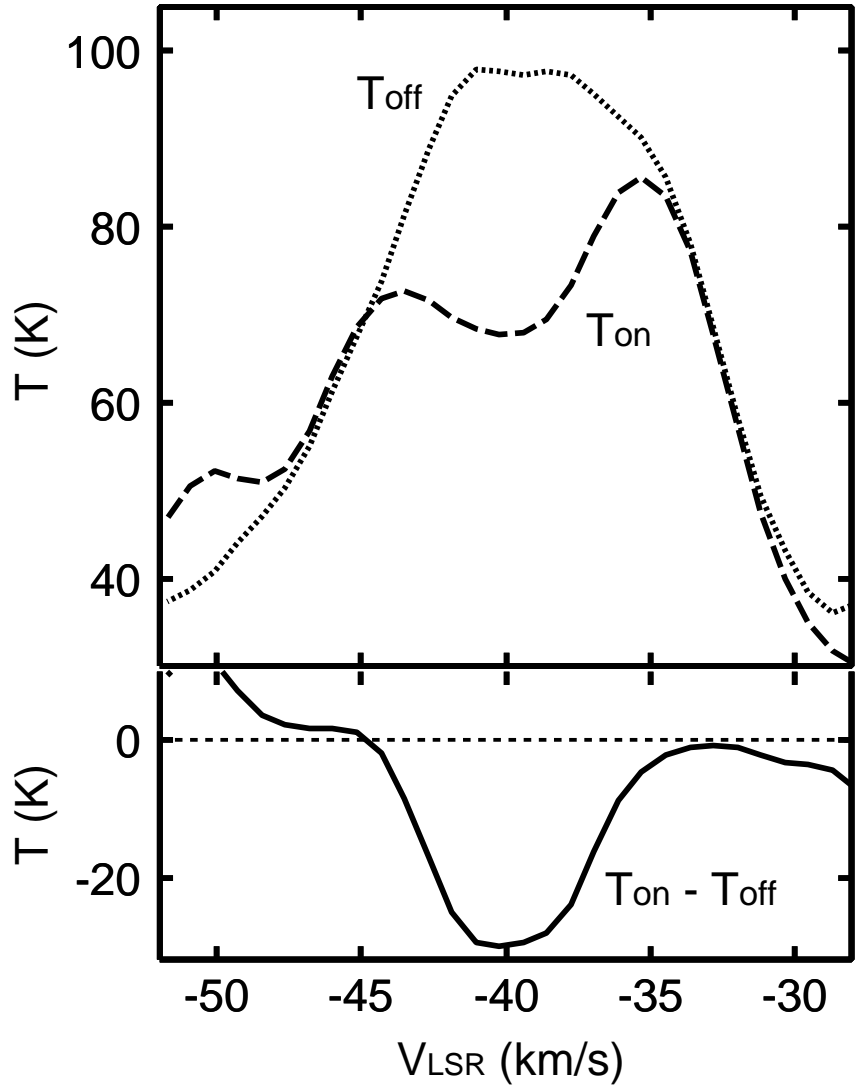}
\caption{
{\it Upper panel} : variation in the brightness temperature with the
velocity, $v_{\rm LSR}$ in the on-region (broken line) and 
off-region (dotted line) in the W5 H~II region.
The positions of the on- and off-region are presented in the lower
left panel in Figure \ref{fig:w5}.
{\it Lower panel} : the absorption line profile obtained by
$T_{\rm on} - T_{\rm off}$. 
}
\label{fig:w5_vprofile}
\end{center}
\end{figure}

\begin{figure}[tb]
\begin{center}
\includegraphics[width=0.8\hsize]{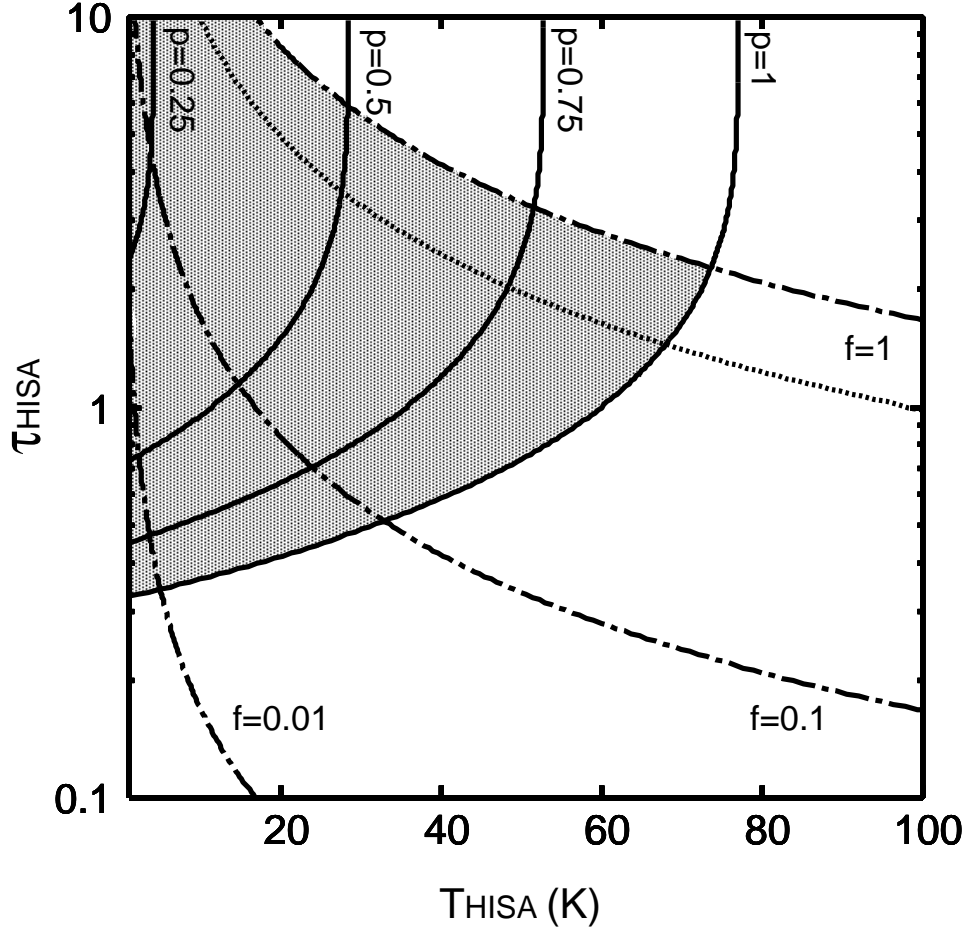}
\caption{ Constraints on the spin temperature and optical depth of the
HISA cloud in the W5 region. The solid curves present the relation
between $T_{\rm HISA}$ and $\tau_{\rm HISA}$ given by equation
(\ref{eq:tsa}) for 
$p~(\equiv \tau_{\rm bg} T_{\rm bg}/T_{\rm off}) = 1.0$, 0.75, 0.5 and 
0.25. The dot-solid curves represent the relation given by equation 
(\ref{eq:dl90}) for $f~(\equiv N_{\rm HI}/N_{\rm H})= 1.0$, 0.1, 0.01 
with $N_{\rm H} = 1.6 \times 10^{21}~{\rm cm}^{-2}$, which is estimated 
from the extinction map.
We also plot the same relation for $f=1.0$ with 
$N_{\rm H} = 9.5 \times 10^{20}~{\rm cm}^{-2}$, which is estimated
from {\it IRAS} $100~\mu$m and $60~\mu$m images (dotted curve).
The shaded region denotes the allowed range of $T_{\rm HISA}$ and
$\tau_{\rm HISA}$ under our constraints.  
}
\label{fig:hisa_w5}
\end{center}
\end{figure}

In order to constrain the physical state of the observed HISA
cloud, we quantitatively analyze the data.
We define ``on''- and ``off''-regions designated 
by $6' \times 6'$ squares in the lower-left panel of Figure \ref{fig:w5}.
The on-region is just on the HISA feature around the W5 H~II region, 
and the off-region is slightly outside the HISA feature.
Figure \ref{fig:w5_vprofile} presents the averaged brightness temperatures
in the on-region ($T_{\rm on}$) and off-region ($T_{\rm off}$) at
different velocities. The emission profile in the on-region clearly shows 
a depression compared with that in the off-region.
We regard this absorption feature as the HISA profile, and obtain the 
line profile calculating $T_{\rm on} - T_{\rm off}$. 
The line depth and width are about $\Delta T \sim 30$~K, 
and $\Delta v \sim 5$~km/s respectively.
The velocity at the line center is $v_{\rm LSR} \sim -40$~km/s, which is
the velocity of the channel maps presented in Figure \ref{fig:w5}.
The HISA profile is available to limit $T_{\rm HISA}$ and 
$\tau_{\rm HISA}$ considering the radiative transfer under a given
geometry of the system \citep[e.g.,][]{Gb00}.
We suppose that the HISA cloud, whose spin temperature and 
optical depth are $T_{\rm HISA}$ and $\tau_{\rm HISA}$, lies
in the sea of the WNM. The diffuse WNM is distributed in both 
the foreground and background of the HISA cloud.
The spin temperature and the optical depth of the foreground
(background) WNM are represented as $T_{\rm fg}$ ($T_{\rm bg}$)
and $\tau_{\rm fg}$ ($\tau_{\rm bg}$).
Note that the spin temperature of the WNM differs from the
kinetic temperature by a factor of two or three, because the thermal
equilibrium is not achieved owing to its low density, 
$n \sim 0.1~{\rm cm}^{-3}$ \citep[e.g.,][]{Lz01}.
The continuum radiation includes the free-free emission from the 
H~II region, cosmic microwave background and so on.
In our case, the W5 H~II region is so diffuse that the brightness
temperature of the continuum radiation, $T_{\rm c}$, is much smaller
than that of the line emission.
Presuming that all the gas components except the HISA cloud are
optically thin, the radiative transfer equation along the line of 
sight in the on- and off-region are,
\begin{equation}
T_{\rm on} = \tau_{\rm fg} T_{\rm fg}
             + T_{\rm HISA} ( 1 - e^{-\tau_{\rm HISA}} )
             + \tau_{\rm bg} T_{\rm bg} e^{- \tau_{\rm HISA}}
             + T_{\rm c} e^{-\tau_{\rm HISA}} - T_{\rm c},
\label{eq:ton}
\end{equation}
\begin{equation}
T_{\rm off} = \tau_{\rm fg} T_{\rm fg} + \tau_{\rm bg} T_{\rm bg} ,
\label{eq:toff}
\end{equation}
where $T_{\rm on}$ ($T_{\rm off}$) is the brightness temperature observed
in the on- (off-) region. Subtracting equation (\ref{eq:toff}) from 
equation (\ref{eq:ton}), we obtain,
\begin{equation}
T_{\rm on} - T_{\rm off} = (1 - e^{- \tau_{\rm HISA}})
                           (T_{\rm HISA} - p T_{\rm off} - T_{\rm c}) 
\label{eq:tsa}
\end{equation}
\citep[e.g.,][]{Gb00, Kv03, Kt05},
where $p$ is the ratio of the background WNM emission to the total
emission, $p \equiv \tau_{\rm bg} T_{\rm bg}/T_{\rm off}$.
We derive the relation between $T_{\rm HISA}$ and $\tau_{\rm HISA}$ 
for a given $p$ from equation (\ref{eq:tsa}), because $T_{\rm on}$, 
$T_{\rm off}$ and $T_{\rm c}$ are all observable quantities. 
We adopt the numerical values of $T_{\rm on} = 67.9$~K, 
$T_{\rm off} = 97.3$~K at the line center, and $T_{\rm c} = 9.1$~K. 
In Figure \ref{fig:hisa_w5}, the solid lines represent the 
$T_{\rm HISA}$-$\tau_{\rm HISA}$ relations 
with these values for a given $p$.
As this figure shows, we can set the maximum spin temperature 
of the HISA cloud as $T_{\rm HISA} \leq 77$~K, where the equality is
satisfied when $p = 1$. Inversely, if the foreground WNM emission
accounts for more than 25\% (50\%) of the total WNM emission, which
corresponds to $p < 0.75$ (0.5), the spin temperature should be 
lower than 52.7~K (28.3~K).
The estimated temperature is comparable to the
temperature predicted by our calculations, and 
{\em much lower than that of the typical PDR}, 
$T \sim 100 - 1000$~K \citep[][Papers I and II]{HT99}. 

We can set further constraint on the $T_{\rm HISA}$-$\tau_{\rm HISA}$ 
plane using the equation
\begin{equation}
\tau_{\rm HISA} = 5.2 \times 10^{-19}
                  \frac{N_{\rm HI}}{\Delta v~T_{\rm HISA}} 
\label{eq:dl90}
\end{equation}
\citep{DL90}, where $N_{\rm HI}$ is the column density of atomic 
hydrogen in the HISA cloud. The line width, $\Delta v$ is about
5~km/s from the line profile presented in Figure \ref{fig:w5_vprofile}.
Therefore, another $T_{\rm HISA}$-$\tau_{\rm HISA}$ relation
is obtained by equation (\ref{eq:dl90}) evaluating $N_{\rm HI}$.
Taking into account that the HISA cloud can be partially  
molecular, the atomic column density is written as 
$N_{\rm HI} = f N_{\rm H}$, where $N_{\rm H}$ is the total column
density of all the hydrogen nuclei and $f$ is the atomic fraction 
in the HISA cloud.
We evaluate the total column density in two different
ways, instead of directly measuring the atomic column density. 
First, we use the extinction map given by \citet{Db05}.
They have analyzed the database of the Digitized Sky Survey I
\citep{Lk94} using the star count technique, and completed an 
extinction map of the entire the Galactic plane. 
We extract the $A_{\rm V}$ map at $6'$ angular resolution around
the W5 region. 
In order to estimate the extinction only through the HISA cloud, 
we calculate the average extinctions in the on- and off-regions, 
and derive their difference. 
The excess extinction in the on-region is $A_{\rm V} = 0.81$, 
which corresponds to $N_{\rm H} \sim 1.6 \times 10^{21}~{\rm cm}^{-2}$. 
With this value of $N_{\rm H}$, we plot
the $T_{\rm HISA}$-$\tau_{\rm HISA}$ relation given by 
equation (\ref{eq:dl90}) for $f=1.0$, 0.1 and 0.01 in 
Figure \ref{fig:hisa_w5}. 
In \S~\ref{ssec:fid} and \ref{ssec:dep_star},
our numerical models have predicted that the gas phase of $f \sim 0.5$ and 
$T \sim$ a few $\times 10$~K should form around the H~II region.
Figure \ref{fig:hisa_w5} shows that such a gas phase is still possible
under our constraints. Second, let us evaluate the total column density 
in another way, using the 60- and $100-{\rm \mu m}$ dust emissions 
\citep[e.g,][]{AG99}. The intensity of the thermal and optically-thin 
dust emission is given by,
\begin{equation}
I_\lambda = \tau_\lambda B_\nu (T_{\rm d}) ,
\label{eq:ilda}
\end{equation}
where $\tau_\lambda$ is the dust opacity, $B_\nu$ is the Planck function,
and $T_{\rm d}$ is the dust thermal temperature. 
The dust opacity is written as,
\begin{equation}
\tau_\lambda = \epsilon_\lambda N_{\rm H} ,
\label{eq:tlda}
\end{equation}
where $\epsilon_\lambda$ is the dust emissivity per hydrogen atom,
which is usually expressed as 
$\epsilon_\lambda = \epsilon_0 \lambda^{- \beta}$.
Although $\beta$ and $\epsilon_0$ actually 
vary with the different environments \citep[e.g.,][]{Dp03}, we adopt 
the typical emissivity law in the local ISM,
\begin{equation}
\epsilon_\lambda = 1.0 \times 10^{-25}
                   \left(  \frac{\lambda}{250~{\rm \mu m}}
                   \right)^{-2} ~ {\rm cm}^2 
\label{eq:elda}
\end{equation}
\citep{Bl96}.
Using equations (\ref{eq:ilda}), (\ref{eq:tlda}) and (\ref{eq:elda}),
we can derive $T_{\rm d}$ and $N_{\rm H}$ from the intensities
at 60 and $100~\mu$m,
$I_{\rm 60 \mu m}$ and $I_{\rm 100 \mu m}$.
We analyze the {\it IRAS} $60-\mu$m and $100-\mu$m images, 
and extract the dust emission only from the dust-shell
by calculating the difference in intensity between on- and
off-regions. The excess intensities in the on-region are 
$I_{\rm 60 \mu m} = 89.1$~MJy/str and
$I_{\rm 100 \mu m} = 180.3$~MJy/str. With these values, we obtain
$T_{\rm d} = 29.5$~K and $N_{\rm H} = 9.5 \times 10^{20}~{\rm cm}^{-2}$.
The derived total column density is smaller than that from
the extinction map by a factor of 1.7 only. In Figure \ref{fig:hisa_w5},
we also plot the $T_{\rm HISA}$-$\tau_{\rm HISA}$ relation given by
equation (\ref{eq:dl90}) with this value of $N_{\rm H}$
for $f = 1.0$. Note that the 60-$\mu$m emission is actually
contaminated with the emission from very small grains which
are not in thermal equilibrium. Therefore, the above values
of $T_{\rm d}$ and $N_{\rm H}$ should be considered as the upper
and lower limits. 
The future high-resolution observation at $\lambda > 100~{\rm \mu m}$ 
will provide tight constraints.

\section{Discussions}
\label{sec:disc}

\subsection{Strength of the FUV Background Field}

In our numerical modeling, we have adopted an FUV background field
of $G_{\rm FUV} = 1.0$ as the fiducial value. 
However, the FUV background field is generally variable with time, 
and sensitive to proximal massive stars \citep{PHM03}.
Quantitatively, a stronger FUV field reduces the molecular
abundance in the shell. This is because the self-shielding
effect becomes inefficient with a stronger FUV background field, 
namely the ratio, $\chi \equiv G_{\rm FUV}/n$ becomes larger with 
larger $G_{\rm FUV}$. 
As expected with the equilibrium $n$-$P$ curves 
(also see \S~\ref{sec:num}), the shell density decreases 
as $G_{\rm FUV}$ increases under the same pressure, and $\chi$
increases. We have examined some models 
with different FUV background fields. With $G_{\rm FUV} = 3.3$ (10) 
in model CNM-S41, for example, only about 25~\% (2~\%) of hydrogen 
atoms exist as H$_2$ molecules in the shell at $t \sim 4.5$~Myr. 
Remember that the H$_2$ ratio is 45~\% in the
same model with $G_{\rm FUV} = 1.0$ (see \S~\ref{ssec:fid}).
The gas temperature in the shell is about 40~K (60~K)
with $G_{\rm FUV} = 3.3$ (10), which is still much lower than 
the typical CNM temperature. 
This situation may be more suitable for HISA clouds 
around the W4 and W5 H~II regions, because the FUV field should 
be higher than the local value in a massive star-forming region.
Inversely, the H$_2$ abundance ratio is slightly enhanced 
with a weaker background field. However, the FUV field 
in the shell does not fall below $G_{\rm FUV} \sim 1$ owing to
the FUV radiation from the central star. CO molecules seldom
accumulate in the shell, and the ``dark'' character of the shell
does not change with a weaker background field.

\subsection{Reformation Rate of Hydrogen Molecules}
 
In previous sections, we have seen that the expanding H~II 
region sometimes triggers the formation of H$_2$ clouds
without CO from the ambient neutral medium. 
These H$_2$ molecules are formed on dust grains in the dense shell, 
and we included the reaction rate given by \citet{HM79} 
in our numerical calculations. However, there is still a large 
uncertainty on the formation rate of H$_2$ molecules. 
It is not surprising that the formation rate depends on the 
local physical state of the ISM.
For example, \citet{Hb04} have suggested that the formation rate
should be about five times enhanced in some nearby PDRs.
Taking the uncertainty into account, we have calculated model 
CNM-S41 with different formation rates of $5 \times R_f$ and 
$1/5 \times R_f$, where $R_f$ is the adopted normal reaction rate. 
Whereas about 45~\% of H atoms are included 
in H$_2$ molecules with $R_f$, H$_2$ fraction becomes 
83~\% (3~\%) with $5 \times R_f$ ($1/5 \times R_f$) at 
$t \sim 4.5$~Myr.
Since the self-shielding is the dominant process to shield the
FUV radiation, the H$_2$ abundance within the shell is somewhat
sensitive to the formation rate. In spite of different H$_2$ 
abundance in the shell, however, only less than 0.1~\% of C 
atoms are converted to CO molecules in all models. 
As mentioned in \S~\ref{ssec:ex_cnm}, the dust absorption of the
FUV radiation with the sufficiently large $N_{\rm sh}$ is needed 
for the accumulation of CO molecules.

\subsection{Expansion of Bubbles and Superbubbles}
\label{ssec:bubble}

Although we have only focused on the expansion of the H~II region in
this paper, the strong stellar wind generally affects the dynamics
around a main-sequence massive star.
When the wind-driven bubble expands in a preexistent H~II region,
the forward SF sweeps up an ionized medium and the 
ionized shell forms around the hot bubble \citep{Wv77}. 
The ionized shell observed in the W4 
H~II region (Figure \ref{fig:w4}) may correspond to this feature.
In the diffuse ambient medium of $n < 10~{\rm cm}^{-3}$, however,
the effect of the wind-driven bubble is not so significant
as for expansion around a single star.
We can easily show that the bubble pressure quickly decreases 
to the initial H~II pressure, and the bubble is confined to the
H~II region \citep{MK84, HI06b}. 
The dynamical effect of the bubble is highly enhanced when
the giant bubble expands around an OB association 
\citep[superbubble; e.g.,][]{THI81, MK87}.
A huge energy input is maintained over several 10~Myr
by the stellar winds and continuous supernova explosions. 
The expansion is mainly driven in the later supernova
phase, and the energy input rate is 
$L_{\rm sb} \sim 6 \times 10^{35} N_*$~erg/s, 
where $N_*$ is the number of OB stars. 
The time evolution of the superbubble size and pressure 
is given by,
\begin{equation}
R_{\rm sb}(t) \sim 376~{\rm pc} 
               \left( \frac{N_*}{100} \right)^{1/5}
               \left( \frac{n}{0.1~{\rm cm}^{-3}} \right)^{-1/5}
               \left( \frac{t}{10~{\rm Myr}} \right)^{3/5} ,
\end{equation}
\begin{equation}
P_{\rm sb}(t) \sim 0.9 \times 10^{-12}~{\rm dynes/cm^2}
               \left( \frac{N_*}{100} \right)^{2/5}
               \left( \frac{n}{0.1~{\rm cm}^{-3}} \right)^{3/5}
               \left( \frac{t}{10~{\rm Myr}} \right)^{-4/5}  
\label{eq:psb}
\end{equation}
\citep{Wv77, MK87}.
In \S~\ref{ssec:ex_wnm}, we have shown that the H~II region causes 
little dynamical effect in the ambient WNM. However, equation 
(\ref{eq:psb}) means that the superbubble pressure can be about ten 
times as high as the WNM pressure with $G_{\rm FUV} \sim 1$ 
(also see Figure \ref{fig:eqil}).
The SF emerges owing to the high bubble pressure, and a
dense shell forms around the superbubble. We can expect a shell 
density with an equilibrium $n$-$P$ curve, as carried out in 
\S~\ref{ssec:ex_cnm}. 
The swept-up WNM should tend forward the equilibrium state determined
by the enhanced pressure and the FUV field at each time step. 
Adopting a post-shock pressure of $10^{-12}~{\rm dynes/cm}^2$
and $G_{\rm FUV} \sim 1$ in the shell, we expect that
the shell density can be as high as several $\times 10^2~{\rm cm}^{-3}$, 
though the Mach number of the SF is quite low. 
The shell temperature should be several $\times$ 10~K. 
The molecular abundance in the shell depends on the
column density of the shell, which is estimated as,
\begin{equation}
N_{\rm sh}(t) \sim \frac{n R_{\rm sb}(t)}{3}
              \sim 4 \times 10^{19}~{\rm cm}^{-2} 
               \left( \frac{N_*}{100} \right)^{1/5}
               \left( \frac{n}{0.1~{\rm cm}^{-3}} \right)^{4/5}
               \left( \frac{t}{10~{\rm Myr}} \right)^{3/5} .
\end{equation}
The expected column density is much lower than that 
of our numerical models presented in \S~\ref{ssec:ex_cnm}.
The situation should be very difficult for molecular formation in the shell. 
Not only CO, but H$_2$ molecules suffer from photodissociation
by the FUV background field. Even with the higher ambient density of 
$n \sim 1~{\rm cm}^{-3}$, the column density is not high enough for the 
accumulation of CO molecules.
The above estimation agrees with the fact that 
{\em galactic supershells are usually detected with the H~I and/or 
dust emission, but seldom associated with molecular emission} 
\citep[e.g.,][]{HB97}.
In some regions, however, the discrete molecular clouds are
clearly distributed along the supershell \citep[e.g.,][]{Fk99, Yg01}.
If these molecular clouds are formed from the swept-up neutral
medium, an extra shielding effect should be necessary to prevent
the photodissociation of CO molecules. Possible processes are 
shell-fragmentation and subsequent contraction of each fragment. 
If each fragment contracts to a dense core, the column density increases 
and CO molecules will be protected against FUV photons.
Both dynamical and chemical evolution of the shell and fragments 
should be studied in detail in a separate paper.

\section{Conclusions}
\label{sec:conc}

In this paper, we have studied the role of the expanding H~II region 
in an ambient neutral medium. 
The time evolution of the H~II region and swept-up shell has
been studied by solving the UV/FUV radiative transfer and
the thermal/chemical processes in a time-dependent hydrodynamics
code. We have analyzed the chemical structure of the shell, and 
examined how efficiently the molecular gas can reform in the shell.
First, we have studied the expanding H~II region in the ambient CNM.
We have analyzed its expansion around a massive star of 
$M_* = 41~M_\odot$ as a fiducial model.
\begin{itemize}
\item[1.] The SF emerges in front of the IF and sweeps up 
the ambient CNM. H$_2$ molecules reform in the shell, but CO molecules 
seldom form. This is due to the different
efficiencies of the self-shielding effects. Even if H$_2$ molecules
accumulate in the shell, CO molecules cannot form without the 
sufficient column density of the shell.
\item[2.] The swept-up shell becomes cold ($T \sim 30~{\rm K}$) and
dense ($n \sim 10^3~{\rm cm}^{-3}$). This physical state
in the shell is the equilibrium of the pressurized neutral
medium with the FUV radiation field in the shell at each time step. 
The gas pressure rises to about the ram pressure in front of the SF, 
just after the gas is engulfed in the shell. 
\item[3.] 
The shell will fragment into small clouds due to gravitational instability.
The typical mass- and length-scale of
each fragment are $\sim 10~M_\odot$ and $\sim 0.1~{\rm pc}$.
If each fragment contracts into a dense core and increases
the column density, CO molecules may form. 
The cooling layer just behind the SF also suffers 
from the thermal instability. This layer will fragment into tiny 
($\sim$ tens of AUs) cloudlets with the translational motion of a few km/s. 
\end{itemize}
The physical and chemical structure of the shell clearly 
depend on the mass, or UV/FUV luminosity of the central star.
The dependences are summarized as follows: 
\begin{itemize}
\item[4.] The formation of H$_2$ molecules is more efficient
for a higher-mass central star.
When the stellar mass is less than $18~M_\odot$, neither CO
nor H$_2$ molecules form in the shell.
\item[5.]
In the model with the higher-mass star, the shell density and
pressure are higher. 
The denser shell becomes gravitationally unstable earlier,
and the thermal instability also works efficiently in
the highly pressurized shell. Therefore, the effect of these 
instabilities is more significant with the higher-mass star.
\end{itemize}
Our numerical models have shown that cold and dense 
HI/H$_2$ gas is forms in the shell without CO molecules, 
and this formation is just the intermediate phase between the neutral 
medium and molecular clouds. In order to probe this ``dark'' hydrogen, 
we have searched for HISA features associated with  
giant H~II regions using the GCPS data.
Our findings are as follows:
\begin{itemize}
\item[6.] We have found shell-like HISA features around the
ionized gas in both the W5 and W4 H~II regions. 
The shell-like HISA shows good spatial correlation with
the dust shell, but not with the $^{12}$CO(1-0) emission.
\item[7.] According to our quantitative analysis of the data,
the observed HISA cloud can be in the cold ($T \sim$ a few $\times 10$~K)
gas phase that includes a significant amount of 
H$_2$ molecules, just as in the numerical simulations.
\end{itemize}
Finally, we have also calculated the time evolution of the H~II region
in the WNM, but the hydrodynamical evolution is much simpler,
owing to the small pressure excess of the photoionized gas.
When the H~II region reaches the WNM, therefore, the UV radiation 
simply transforms WNM into the WIM.

{\acknowledgements
We are grateful to Yasuo Fukui, Anthony Whitworth, Toshikazu Onishi
and Akiko Kawamura for the fruitful discussions and comments.
The numerical calculations were carried out on Altix3700 BX2
at YITP in Kyoto University. The research presented in this paper 
has used data from the Canadian Galactic Plane Survey, a Canadian 
project with international partners, supported by the Natural Sciences 
and Engineering Research Council. 
We have also analyzed the data obtained from the Atlas and Catalog of 
Dark Clouds, supported by Tokyo Gakugeki University.
SI is supported by the Grant-in-Aid  (15740118, 16077202)
from the Ministry of Education, Culture, Sports, Science, and
Technology (MEXT) of Japan.}

\clearpage
\end{document}